\documentclass[sigconf]{acmart}
\renewcommand\footnotetextcopyrightpermission[1]{}

\usepackage{adjustbox}
\usepackage{multirow}
\usepackage{pifont}
\usepackage{colortbl}
\usepackage{tabularx}
\usepackage{enumitem}
\usepackage{tcolorbox}
\usepackage{pdfpages}

\setcopyright{none}

\begin{document}

\settopmatter{printacmref=false}

\newcommand{\stitle}[1]{\vspace*{0.5em}\noindent{\bf #1\/}}

\title{Rethinking Positional Encoding and Interaction Granularity for CTR Prediction}

\author{{\Large Xinxun Zhang $^{1}$, Yuxiang Wang $^1$, Xiaokai Zhou $^1$, Jiayue Li $^1$, Yi Li $^2$, Yue Liu $^2$, Jun Zhang $^{2\dagger}$, Xiaoling Zhou $^3$, Tieyun Qian $^1$, Xiao Yan $^1$, Jiawei Jiang $^{1\dagger}$}}

\affiliation{
  \institution{$^1$ Wuhan University, China; $^2$ Tencent Inc., China; $^3$ Peking University, China}
  \country{}
}
\email{\{xxzhangstu, nai.yxwang, xiaokaizhou, jiayue.li36, qty, yanxiaosunny, jiawei.jiang\}@whu.edu.cn}
\email{\{sincereli, herculesliu, neoxzhang\}@tencent.com, xiaolingzhou@stu.pku.edu.cn}

\renewcommand{\shortauthors}{Zhang et al.}

\begin{abstract}
Click-through rate (CTR) prediction relies on historical behaviors and contextual features to estimate clicks on a target item. Existing Transformer-based CTR models often adopt index-based or temporal positional encodings and token-level causal interaction. However, we observe that most existing positional encodings do not outperform NoPE, while relaxing causal constraints within target-relative temporal groups also yields gains. These observations motivate a fundamental question: what positional encoding and interaction granularity does CTR prediction actually require? To address this question, we propose \textbf{TRIP}, which combines target-aware rank--time positional encoding with temporal-group interaction modeling. Specifically, Factorized Rank Encoding represents each behavior's ordinal position within the historical sequence through a low-rank factorization, while Target-Relative Time Encoding characterizes its temporal distance to the target on a coarse-to-fine scale. This joint design grounds each behavior's positional representation in its sequence context and temporal relation to the current target. Furthermore, Temporal Group-Causal Interaction enriches intra-group representations through bidirectional interaction and masked reconstruction while preserving inter-group causal order. Finally, behavior representations are condensed into compact group summaries for efficient group-level modeling in subsequent layers. Experiments on four public benchmarks show that TRIP improves AUC by up to 4.45\% relative to the strongest baseline.
\end{abstract}

\keywords{Recommendation System, Click-Through Rate Prediction, Behavior Sequence Modeling}

\maketitle
\begingroup
\renewcommand{\thefootnote}{\ensuremath{\dagger}}
\footnotetext{Corresponding authors.}
\endgroup

\section{Introduction}

Click-through rate (CTR) prediction estimates the probability that a user clicks a target item based on historical behaviors, user and item features, and contextual information~\cite{zhou2018din,song2019autoint}. By modeling heterogeneous feature interactions and historical behavior sequences, CTR prediction provides engagement estimates that guide candidate ranking and exposure~\cite{song2019autoint,zhou2018din,wang2021dcnv2}. CTR models are widely used in e-commerce recommendation~\cite{gu2020dmt}, personalized content delivery~\cite{wu2019npa}, and online advertising~\cite{zhou2018din}. These applications support personalized information access and discovery on the Web~\cite{das2007googlenews,wu2019npa}.

Transformer-based modeling has achieved substantial progress and become an important paradigm for CTR prediction~\cite{song2019autoint,guan2026stca,zhang2026onetrans,cao2026onetransv2,chen2026rest}. Recent approaches unify feature interaction and behavior-sequence modeling within shared Transformer architectures~\cite{zhang2026onetrans,huang2026hyformer,huang2026mixformer,zhou2026tokenformer,liu2026est}. Current methods typically inherit index-based positional encodings and token-level causal attention from language modeling~\cite{zhang2026onetrans,zhou2026tokenformer,chen2026rest}. Positional encodings represent behaviors by their physical indices in the input sequence, while causal attention restricts historical interactions to preceding context. For example, OneTrans~\cite{zhang2026onetrans} processes historical behaviors and non-sequential features as a unified token sequence with learned absolute positional embeddings (APE) through stacked causal Transformer blocks. However, the suitability of these positional encodings and interaction granularities for CTR prediction remains unclear.

\begin{figure}[t]
  \centering
  \begin{minipage}[b]{0.49\linewidth}
    \centering
    \includegraphics[width=\linewidth]{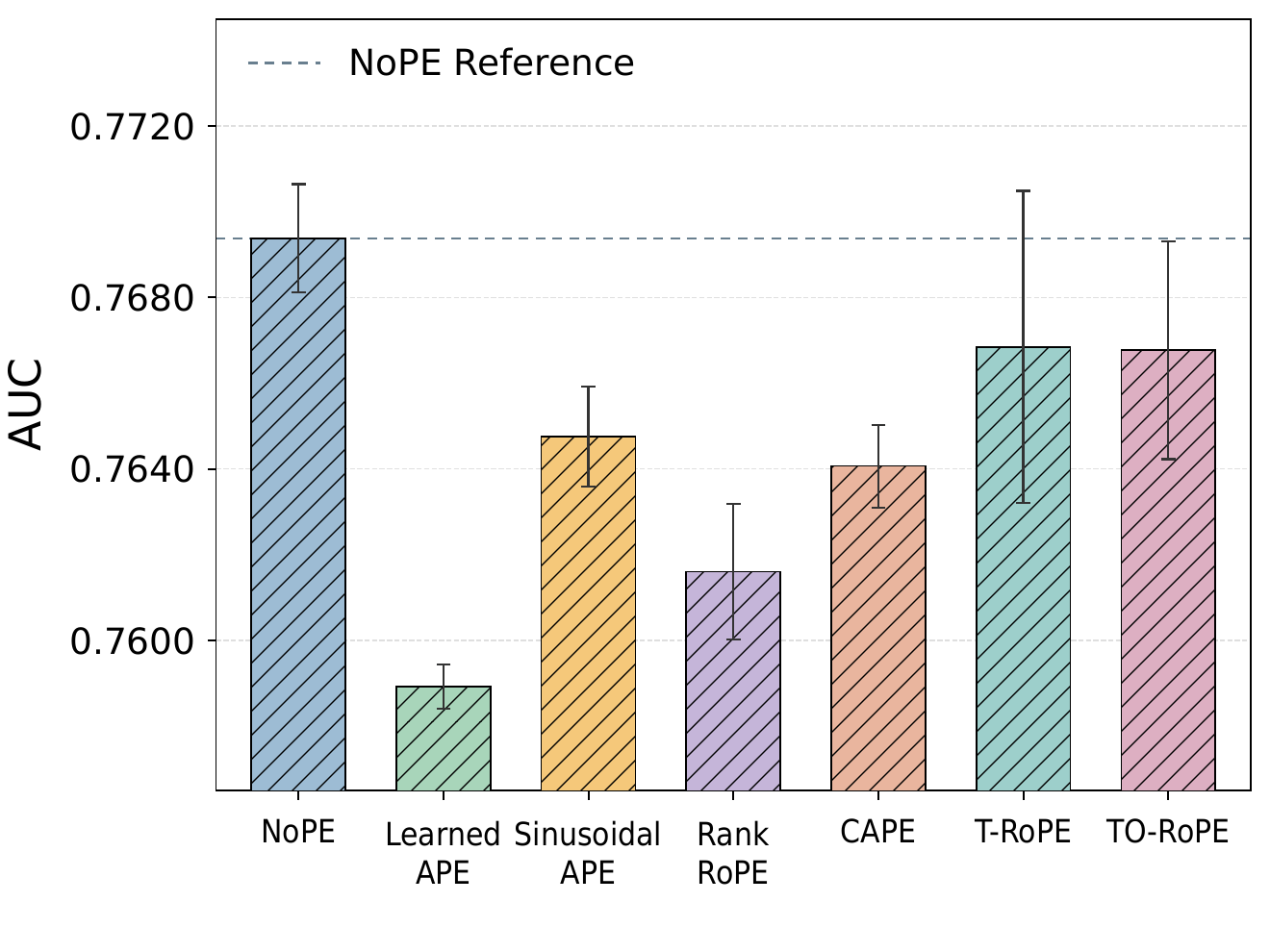}
    \centerline{\small (a) Encoding comparison}
  \end{minipage}\hfill
  \begin{minipage}[b]{0.49\linewidth}
    \centering
    \includegraphics[width=\linewidth]{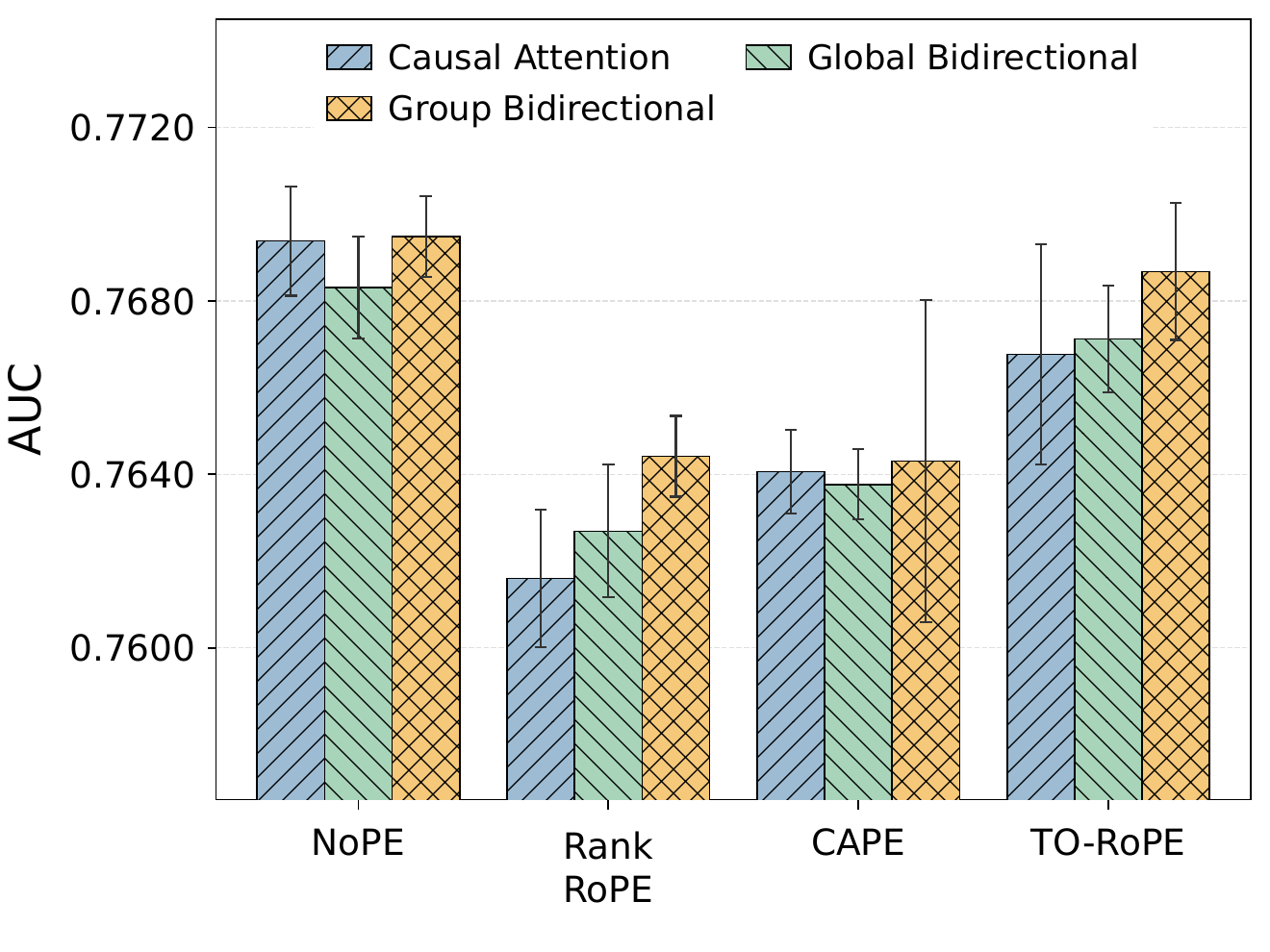}
    \centerline{\small (b) Interaction granularity}
  \end{minipage}
  \caption{Performance of OneTrans on KuaiRand under different positional encodings and interaction granularities. (a) NoPE versus index-based encodings (learned APE, sinusoidal APE, and rank RoPE~\cite{su2024roformer}), time-based encoding (T-RoPE~\cite{liu2026trope}), joint index--time encoding (TO-RoPE~\cite{wei2026torope}), and target-aware encoding (CAPE~\cite{yuan2025cape}). (b) Causal, target-relative temporal-group bidirectional, and global bidirectional attention under NoPE, rank RoPE, CAPE, and TO-RoPE.}
  \Description{Panel a compares NoPE, learned APE, sinusoidal APE, rank RoPE, CAPE, T-RoPE, and TO-RoPE. All six explicit encodings have lower mean AUC than NoPE. Panel b compares causal, group bidirectional, and global bidirectional attention under NoPE, rank RoPE, CAPE, and TO-RoPE. Group bidirectional attention has the highest mean AUC in each displayed configuration. Error bars indicate variation across seeds.}
  \label{fig:motivation}
\end{figure}

\stitle{Empirical Observations.}
We examine positional encoding and interaction granularity in OneTrans~\cite{zhang2026onetrans} and observe two phenomena. (1) As shown in Figure~\ref{fig:motivation}(a), none of the six positional encoding configurations of OneTrans yields better performance than its NoPE counterpart on KuaiRand. (2) Figure~\ref{fig:motivation}(b) shows that bidirectional interaction within target-relative temporal groups outperforms both event-level causal and global bidirectional interaction in AUC under all four encodings.
Together, these findings motivate reconsidering: \emph{What positional information should CTR models encode, and at what granularity should historical behaviors interact?}

\begin{figure}[t]
  \centering
  \includegraphics[width=\linewidth]{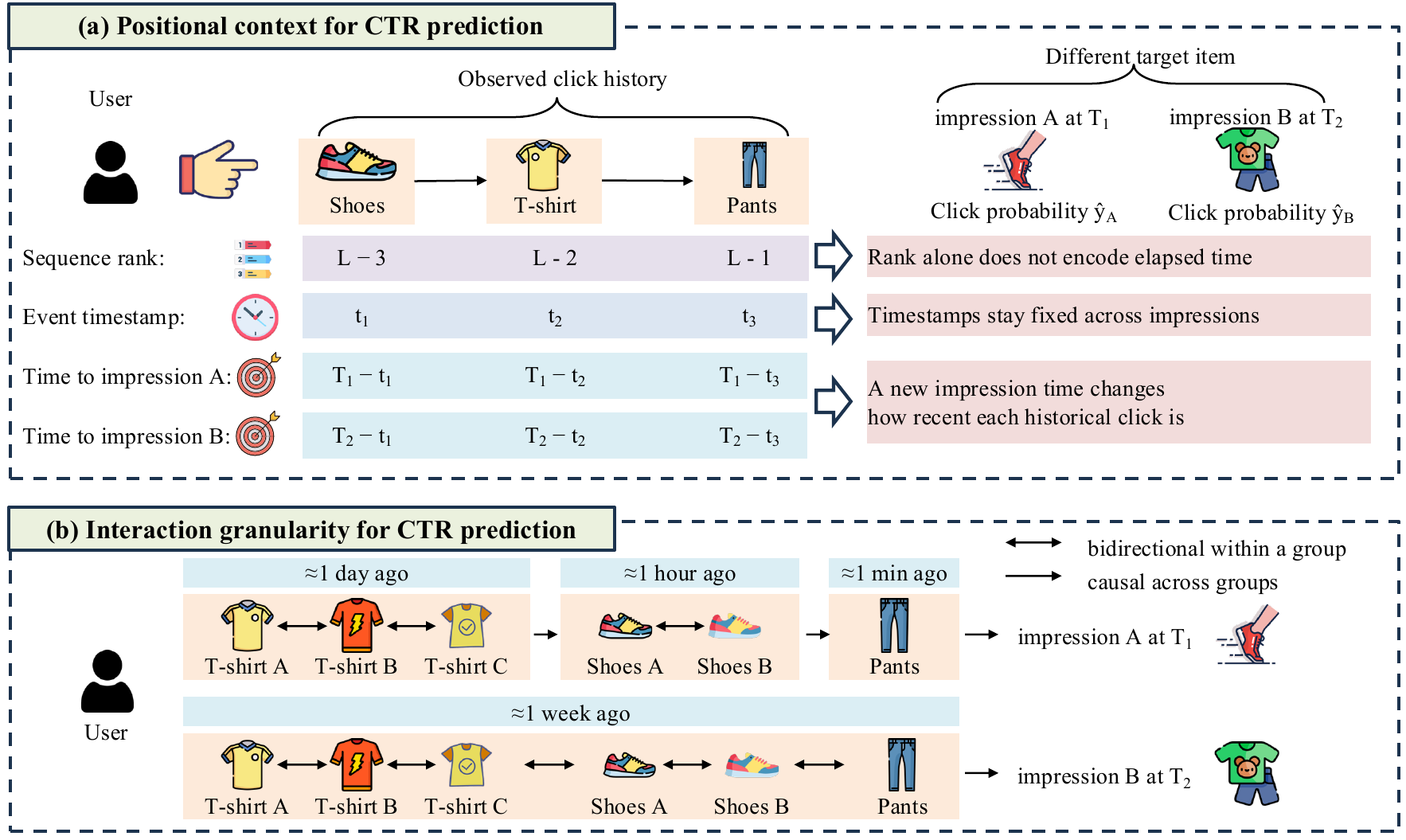}
  \caption{An illustrative example of positional information and interaction granularity in CTR histories. (a) Sequence rank, event timestamps $t_i$, and target-relative time $T_k-t_i$, where $T_k$ is a prediction timestamp. (b) Target-relative temporal groups, with bidirectional interaction within groups and causal interaction across groups.}
  \Description{Panel a shows one history of shoes, a T-shirt, and pants. It distinguishes right-aligned ranks, original event timestamps, and elapsed times to targets predicted at T1 and T2. The unchanged history has different temporal distances when the prediction timestamp changes. Panel b shows T-shirts, shoes, and pants browsed about one day, one hour, and one minute before impression A, forming three temporal groups that widen with distance; groups interact bidirectionally inside and causally across. Viewed from impression B a week later, the same six behaviors form a single group.}
  \label{fig:shopping-motivation}
\end{figure}

\stitle{Understanding the Observations.}
To understand these observations, we examine historical behavior sequences in the context of the current prediction target. Figure~\ref{fig:shopping-motivation} provides an illustrative example for the following two-part analysis.
\noindent\begin{itemize}[leftmargin=*]
  \item \textbf{Positional information.} As illustrated in Figure~\ref{fig:shopping-motivation}(a), historical behaviors in CTR prediction have three positional views: sequence rank, event timestamp, and target-relative time. Sequence rank characterizes a behavior's ordinal context, whereas its timestamp specifies when it occurred. For a fixed history, ranks and timestamps remain unchanged across prediction requests, but the target-relative coordinates change from $T_1-t_i$ to $T_2-t_i$ with the prediction timestamp. Moreover, the same rank can correspond to different target-relative times, while temporally adjacent behaviors can occupy distinct ordinal positions. Thus, effective CTR positional modeling should jointly capture behavioral order, temporal information, and target-relative context, rather than relying on any single positional signal alone.

 \item \textbf{Interaction granularity.} As illustrated in Figure~\ref{fig:shopping-motivation}(b), the same history is partitioned into different temporal groups as the prediction time changes, and is thus modeled at different scales for different targets. Within a group, behaviors mutually enrich one another to characterize the user's interest during that period, while successive groups reveal how interests evolve toward the current request. Event-level causal interaction limits this within-group enrichment, whereas global bidirectional interaction removes the directed order across groups.

\end{itemize}

\stitle{Our Solution.}
Based on the above observations and analysis, we propose \textbf{TRIP}, which combines target-aware rank--time positional encoding with temporal-group interaction modeling. \textbf{Factorized Rank Encoding (FRE)} represents each behavior's ordinal position through a low-rank factorization with rank-specific coefficients over shared basis vectors. This preserves rank distinctions with a lightweight parameterization. \textbf{Target-Relative Time Encoding (TRTE)} encodes each behavior's temporal distance to the target on a coarse-to-fine scale. Based on these target-anchored coordinates, it constructs density-corrected learnable temporal potentials as attention biases to guide time-aware history reading. Using the same target-relative temporal reference, \textbf{Temporal Group-Causal Interaction (TGCI)} partitions the history into temporal groups, where behaviors enrich one another through bidirectional interaction, while causal interaction across groups preserves interest evolution toward the current request. Furthermore, we use masked item reconstruction to strengthen contextual representations within each temporal group. Finally, contextualized behavior states are condensed into compact group summaries for efficient modeling of interest evolution in subsequent layers.

We evaluate TRIP against eleven baselines on four public CTR benchmarks. TRIP outperforms the best-performing baseline on each dataset, achieving an average relative AUC improvement of 1.47\% and a maximum of 4.45\% on TAAC. The joint target aware rank--time encoding also transfers to other Transformer-based CTR models, achieving relative AUC improvements of up to 4.20\% over the best of eight alternative positional encoding configurations on the same backbone. Ablation studies further confirm the importance of behavior rank and target-relative time, as well as the effectiveness of temporal-group interaction modeling. The main contributions of this work are as follows:

\begin{itemize}[leftmargin=*]
  \item We identify limitations of existing positional and temporal encodings and uniform interaction structures in CTR prediction through empirical comparisons and behavioral analysis.
  \item We propose target-aware rank--time positional encoding through FRE and TRTE, combining lightweight ordinal representations with target-relative temporal potentials.
  \item We develop TGCI for within-group mutual enrichment and cross-group causal interest evolution, with masked reconstruction for contextual learning and group summaries for efficient modeling.

\end{itemize}

\section{Preliminaries}
\label{sec:preliminaries}

CTR prediction is a supervised binary classification task that estimates the probability of a click on a displayed target item. Each sample corresponds to an impression of item $v_*$ to user $u$ at prediction time $T$. Let $\mathbf{x}^{\mathrm{NS}}=[x^{(1)},\ldots,x^{(F)}]$ denote the associated user, target-item, and contextual features across $F$ fields. These heterogeneous fields include categorical identifiers and attributes, as well as numerical features. The user's observed behavior history is
\begin{equation}
  \mathcal H_u(T)=\bigl[(v_i,\mathbf f_i,t_i)\bigr]_{i=1}^{n},
  \qquad t_1\le\cdots\le t_n<T,
  \label{eq:ctr-history}
\end{equation}
where $n$ is the history length, and $v_i$, $\mathbf f_i$, and $t_i$ denote the item, associated features, and timestamp of the $i$-th event, respectively. Historical behaviors provide sequential evidence of user interests, while the target item and impression context specify the prediction instance. The label $y\in\{0,1\}$ indicates whether the displayed item is clicked. Conditioned on these inputs, the model estimates
\begin{equation}
  \hat y=f_\theta\bigl(\mathcal H_u(T),v_*,\mathbf{x}^{\mathrm{NS}},T\bigr).
  \label{eq:ctr-prediction}
\end{equation}
The model parameters $\theta$ are learned from labeled interactions by minimizing the binary cross-entropy loss.
Following unified CTR models~\cite{zhang2026onetrans,huang2026hyformer,huang2026mixformer}, we map each historical event to a token $\mathbf h_i=\operatorname{Tokenizer}(v_i,\mathbf f_i,t_i)$ and tokenize the embedded user, target-item, and contextual fields into $N_{\mathrm{ns}}$ non-sequential tokens. The resulting matrices are $\mathbf H_{\mathrm S}\in\mathbb R^{L\times D}$ and $\mathbf Z_{\mathrm{NS}}\in\mathbb R^{N_{\mathrm{ns}}\times D}$, where $L$ is the history capacity and $D$ the hidden dimension. Non-sequential tokens can combine multiple fields, so $N_{\mathrm{ns}}$ need not equal the field count $F$.

\section{Methodology}
\label{sec:methodology}

\begin{figure*}[t]
  \centering
 \includegraphics[width=0.95\textwidth]{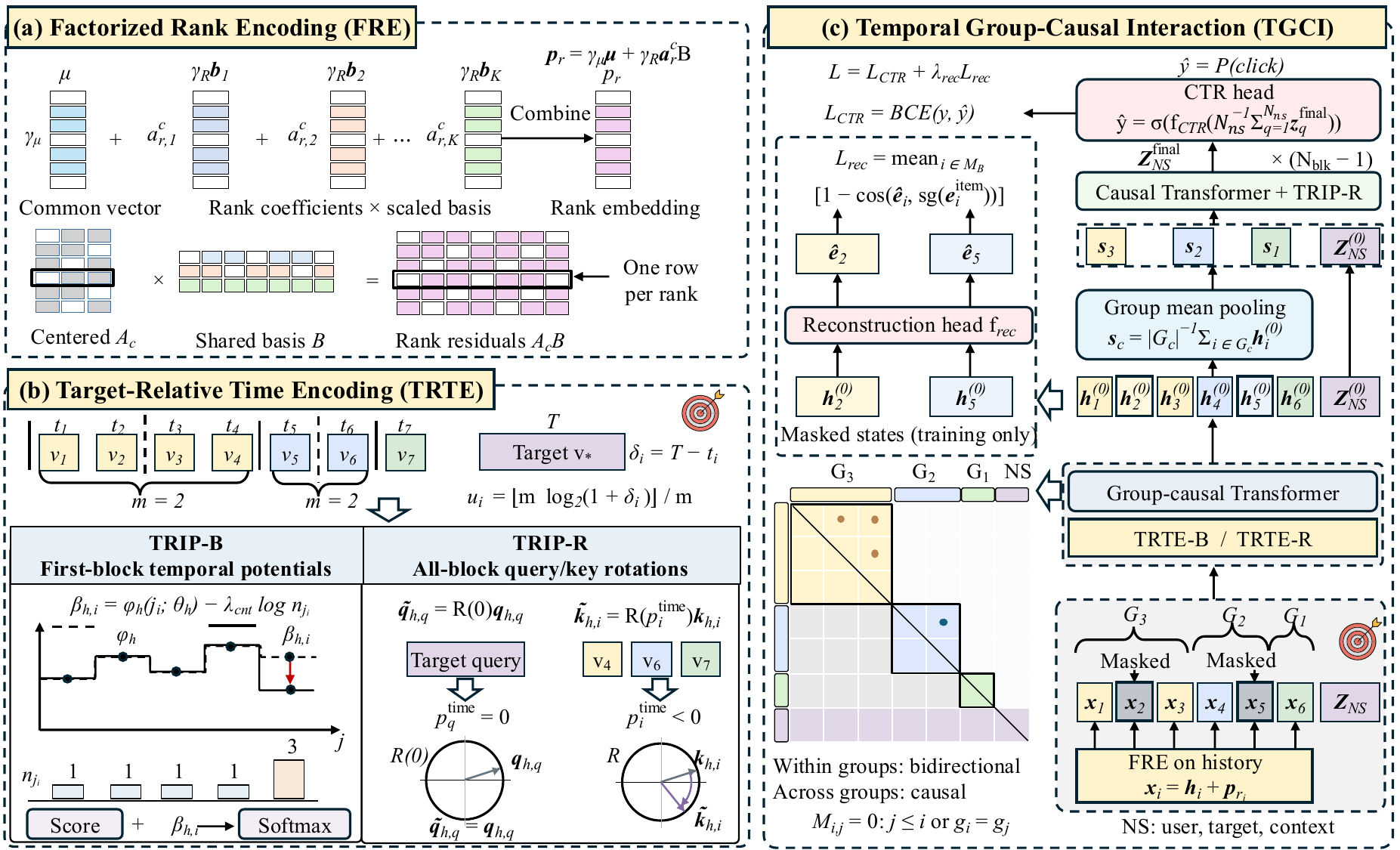}
  \caption{Overview of TRIP. FRE and TRTE encode sequence rank and target-relative time (learnable temporal potentials in TRIP-B or temporal rotations in TRIP-R). TGCI performs intra-group bidirectional and inter-group causal interaction, with masked reconstruction and group-summary compression for subsequent CTR modeling.}
  \Description{The left panels illustrate the low-rank construction of rank embeddings and the two variants of target-relative temporal encoding. The right panel shows the model pipeline, including masking, group-causal attention, reconstruction, group summaries, subsequent causal Transformer blocks, and the CTR head.}
  \label{fig:model}
\end{figure*}
Figure~\ref{fig:model} illustrates TRIP, a unified CTR framework for joint behavior-sequence modeling and non-sequential feature interaction.
TRIP jointly represents behavioral rank and target-relative time, then organizes interactions within target-relative temporal groups. FRE injects a lightweight ordinal representation into each history token. TRTE uses the event's distance to the target to guide attention through either learnable temporal potentials or temporal rotations. TGCI contextualizes events within temporal groups, supplies masked-item reconstruction supervision, and compresses the history into group summaries for subsequent modeling.
\subsection{Factorized Rank Encoding}
\label{sec:fre}

\begin{figure}[t]
  \centering
  \includegraphics[width=\linewidth]{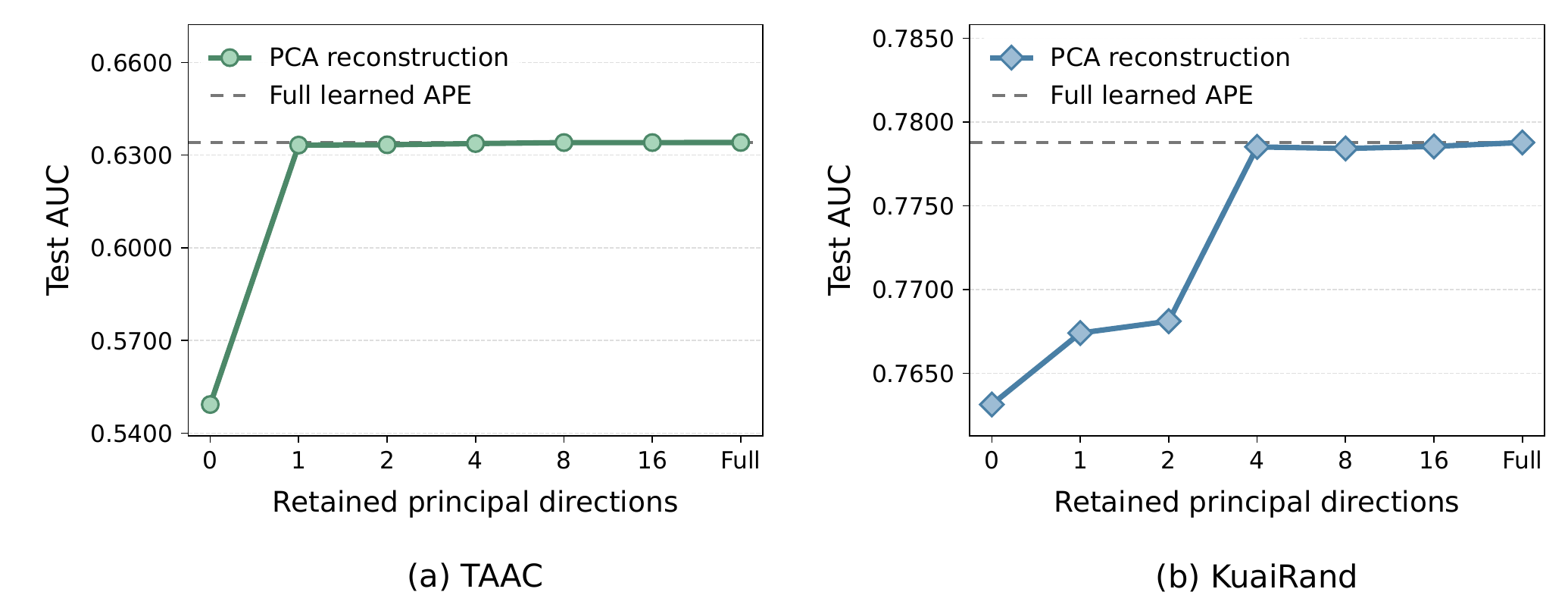}
  \caption{Performance after PCA reconstruction of learned history position embeddings on TAAC and KuaiRand. Full learned APE is the uncompressed history-only position table.}
  \Description{Test AUC is plotted against the number of retained PCA directions. Reconstruction using one direction on TAAC and four on KuaiRand approaches the performance of the full learned historical position table. The analyzed position embeddings are applied only to history tokens.}
  \label{fig:fre-pca}
\end{figure}

\stitle{Observations.} In CTR prediction, sequence rank provides an item's ordinal context, which elapsed time alone cannot fully characterize under irregular user activity. Since the predictive role of a historical behavior can vary across sequence positions, we adopt learnable rank embeddings that allow CTR supervision to determine position-specific representations without imposing a predefined positional geometry. Interestingly, we observe that these learned embeddings exhibit an approximately low-rank structure. As shown in Figure~\ref{fig:fre-pca}, reconstructing them with only a few principal directions largely preserves prediction performance. Motivated by this observation, we design FRE with rank-specific coefficients over shared basis vectors, enabling flexible yet compact ordinal representations.

FRE parameterizes the rank representation with a common vector $\boldsymbol\mu\in\mathbb R^D$, position-specific coefficients $A\in\mathbb R^{L\times K}$, and shared basis vectors $B\in\mathbb R^{K\times D}$. Here, $K$ is the factorization width. We center the coefficients across ranks to separate the shared offset from rank-specific variation and define the rank representation as
\begin{equation}
  A_c=A-\frac{1}{L}\mathbf1\mathbf1^{\top}A,\qquad
  \mathbf p_r=\gamma_\mu\boldsymbol\mu+\gamma_R\mathbf a_r^cB.
  \label{eq:fre-encoding}
\end{equation}
where $\mathbf a_r^c$ is row $r$ of $A_c$, and $\gamma_\mu$ and $\gamma_R$ control the shared and rank-dependent components, respectively. The parameters $\boldsymbol\mu$, $A$, and $B$ are jointly learned end-to-end, allowing each rank to learn a distinct linear combination of shared basis vectors for CTR prediction.

For event $i$ with sequence rank $r_i$, FRE adds the corresponding rank embedding to its input representation:
\begin{equation}
  \mathbf x_i=\mathbf h_i+\mathbf p_{r_i},\qquad
  \mathbf X_{\mathrm{in}}=[\mathbf H_{\mathrm S}+\mathbf P;\mathbf Z_{\mathrm{NS}}],
  \label{eq:fre-input}
\end{equation}
where $\mathbf P$ collects the rank embeddings, and concatenation is along the token axis. To represent ordinal context within the behavior sequence, we only apply FRE to the history tokens. Compared with full learned APE, FRE reduces the positional parameter count from $LD$ to $D+LK+KD$, corresponding to a $3.8$--$21.6\times$ reduction under our benchmark configurations. This lightweight design provides effective rank representations within the behavior history.

\subsection{Target-Relative Time Encoding}
\label{sec:trte}

Sequence rank captures behavioral order but overlooks temporal intervals, while event timestamps alone do not explicitly characterize temporal relationships to the current prediction target. Moreover, the temporal relevance of a historical behavior varies with the current target, even when the history remains unchanged. Therefore, we propose TRTE to encode each behavior's temporal position relative to the target.

\stitle{Target-Anchored Temporal Coordinates.}
Temporal distinctions are not equally informative across the history: recent behaviors often reflect immediate interests and benefit from finer temporal resolution, whereas distant behaviors can be modeled at coarser scales. We therefore adopt a fine-grained logarithmic time scale to preserve detailed temporal distinctions near the target while compactly representing the distant past. The temporal coordinate is defined as
\begin{equation}
  u_i=\frac{\lfloor m\log_2(1+\delta_i)\rfloor}{m},
  \label{eq:trte-scale}
\end{equation}
where $\delta_i=T-t_i$ is the elapsed time in seconds from event $i$ to the prediction timestamp $T$, $m\ge1$ specifies the number of subdivisions per logarithmic interval, and $u_i$ is the resulting discretized logarithmic time coordinate. Increasing $m$ refines temporal resolution while preserving the logarithmic scaling.

\stitle{Learnable Temporal Potentials.}
We map each target-relative coordinate $u_i$ to a discrete temporal interval index $j_i$ and introduce a head-specific learnable function $\phi_h(j_i;\theta_h)$. This allows different attention heads to respond differently to the same temporal interval. Combined with the fine-grained logarithmic time scale, these head-specific preferences enable flexible aggregation of behaviors from both recent and distant history. Historical behaviors are unevenly distributed across temporal intervals. Since softmax assigns weights to individual events, intervals containing more behaviors may accumulate greater attention mass simply through event multiplicity. We therefore correct the temporal potential using the event distribution within each sample. For positive elapsed times, the corrected potential is
\begin{equation}
  \beta_{h,i}=\phi_h(j_i;\theta_h)
             -\lambda_{\mathrm{cnt}}\log n_{j_i},
  \label{eq:trte-potential}
\end{equation}
where $n_{j_i}$ counts valid history events in interval $j_i$ for the current sample, and $\lambda_{\mathrm{cnt}}$ controls the correction strength. This correction mitigates quantity-driven amplification while retaining content relevance and learned temporal preferences.

For prediction-side queries $q\in\mathrm{NS}$ and history keys $k\in\mathrm S$, we add the corrected potential to the attention scores:
\begin{equation}
  \ell_{h,q,k}=\frac{\langle\mathbf q_{h,q},\mathbf k_{h,k}\rangle}{\sqrt{d_h}}+\beta_{h,k}.
  \label{eq:trte-reading}
\end{equation}
Here, $\mathbf q_{h,q}$ and $\mathbf k_{h,k}$ denote query and key vectors, $\langle\cdot,\cdot\rangle$ is their inner product, and $d_h=D/H$ is the dimension of each of $H$ heads. Other query--key pairs retain the standard content-based scores. The attention weights are obtained by softmax:
\begin{equation}
  \alpha_{h,q,k}=\frac{\exp(\ell_{h,q,k})}{\sum_{k'\in\mathcal V(q)}\exp(\ell_{h,q,k'})},
  \label{eq:trte-softmax}
\end{equation}
where $\mathcal V(q)$ contains all valid keys visible to query $q$, including permitted prediction-side keys. The corrected scores guide history reading through head-specific target-relative temporal preferences. This learned-potential variant is denoted TRIP-B.

\stitle{Target-Relative Temporal Rotations.}
As an alternative to learnable temporal potentials, we introduce a rotary formulation, denoted TRIP-R, that incorporates target-relative temporal information into query/key interactions. Reusing the logarithmic coordinate in Equation~\ref{eq:trte-scale}, we place historical behaviors with $\delta_i>0$ on the negative time axis and prediction-side tokens at the origin:
\begin{equation}
  p_i^{\mathrm{time}}=-\min(C_{\mathrm{rot}},1+u_i),\qquad
  p_q^{\mathrm{time}}=0\quad(q\in\mathrm{NS}).
  \label{eq:trte-rotary-coordinate}
\end{equation}
Here, $u_i$ is evaluated with resolution $m=m_{\mathrm R}$, and $C_{\mathrm{rot}}$ bounds the coordinate magnitude. This arrangement expresses each historical behavior's temporal position relative to the current prediction time. Following RoPE~\cite{su2024roformer}, we rotate queries and keys:
\begin{equation}
  \widetilde{\mathbf q}_{h,i}=R(p_i^{\mathrm{time}})\mathbf q_{h,i},\qquad
  \widetilde{\mathbf k}_{h,i}=R(p_i^{\mathrm{time}})\mathbf k_{h,i},
  \label{eq:trte-rotary-qk}
\end{equation}
where $R(\cdot)$ denotes the rotary transformation. The resulting attention scores are
\begin{equation}
  \ell_{h,q,k}=\frac{\langle\widetilde{\mathbf q}_{h,q},\widetilde{\mathbf k}_{h,k}\rangle}{\sqrt{d_h}},
  \label{eq:trte-rotary-score}
\end{equation}
and are normalized using Equation~\ref{eq:trte-softmax}. This formulation encodes target-relative temporal relationships through query/key rotations, while FRE retains ordinal context in the input representations. TRIP-B and TRIP-R are alternative variants of TRTE. TRIP-B biases prediction-side history reading only in the first Transformer block, while TRIP-R applies temporal Q/K rotations in all Transformer blocks. Implementation details are provided in Appendix~\ref{app:module-details}.

\subsection{Temporal Group-Causal Interaction}
\label{sec:tgci}

TGCI organizes the history into temporal groups defined by distance to the current prediction, so that behaviors within a group exchange context bidirectionally while information across groups flows causally toward the target. Sharing the target-relative reference of TRTE with its own resolution, we partition historical behaviors as
\begin{equation}
  g_i=\min\bigl(C_{\mathrm{grp}},\,1+\lfloor\log_2(1+\delta_i)\rfloor\bigr),
  \label{eq:tgci-group-index}
\end{equation}
where $C_{\mathrm{grp}}$ caps the group index and $\mathcal G_c=\{i:g_i=c\}$ denotes a nonempty temporal group. For valid history events ordered chronologically, we define the additive attention mask as
\begin{equation}
  \mathsf M_{i,j}=\begin{cases}
    0,& j\le i\ \text{or}\ g_i=g_j,\\
    -\infty,& \text{otherwise}.
  \end{cases}
  \label{eq:tgci-visibility}
\end{equation}
The mask is added to the attention scores before softmax. It permits bidirectional context exchange within a group while retaining causal information flow across groups. Because groups are defined relative to $T$, the same history is regrouped as the target changes.

\stitle{Masked Contextual Reconstruction.}
To encourage learning semantic relationships within temporal groups, we introduce masked-item reconstruction as auxiliary supervision. During training, we select historical events for masking at a nominal ratio $\rho\in(0,1)$ and replace their item embeddings with a learnable mask embedding. The masked set $\mathcal M$ is selected within temporal groups. We reconstruct the original item embeddings from contextualized event states:
\begin{equation}
  \widehat{\mathbf e}_i=f_{\mathrm{rec}}\bigl(\mathbf h_i^{(0)}\bigr),\qquad i\in\mathcal M,
  \label{eq:tgci-reconstruction}
\end{equation}
where $\mathcal M$ is the set of masked events and $\mathbf h_i^{(0)}$ is the event representation after the first Transformer block. The reconstruction head $f_{\mathrm{rec}}$ consists of layer normalization and a two-layer MLP with a SiLU activation. It predicts the original item embedding from each masked event's contextualized state before group compression. The auxiliary objective encourages recovery of masked behavior content using local context. Reconstruction and CTR prediction share the same masked input and backbone pass during training. At inference, masking and the reconstruction head are disabled.

\stitle{Group-Level Interest Modeling.}
After within-group contextualization, we aggregate event representations into a group summary:
\begin{equation}
  \mathbf s_c=\frac{1}{|\mathcal G_c|}\sum_{i\in\mathcal G_c}\mathbf h_i^{(0)}.
  \label{eq:tgci-summary}
\end{equation}
Each summary represents the user's interest during the corresponding period. The summaries are arranged chronologically and replace the original events, reducing the history length from $L$ events to $G$ nonempty group summaries, while prediction-side states are retained. Subsequent causal Transformer blocks model interest evolution across these group-level representations more effectively and at lower computational cost.

The CTR head predicts the click probability from the mean-pooled prediction-side states:
\begin{equation}
  \hat y=\sigma\left(f_{\mathrm{CTR}}\left(\frac{1}{N_{\mathrm{ns}}}\sum_{q=1}^{N_{\mathrm{ns}}}\mathbf z_q^{\mathrm{final}}\right)\right),
  \label{eq:trip-readout}
\end{equation}
where $\mathbf z_q^{\mathrm{final}}$ is the final state of prediction-side token $q$, $f_{\mathrm{CTR}}$ is an MLP, and $\sigma$ is the sigmoid function.

\subsection{Optimization}
\label{sec:optimization}

We train TRIP by jointly minimizing the CTR prediction loss and an auxiliary masked-item reconstruction loss. For a minibatch $\mathcal B$, the CTR objective is binary cross-entropy:
\begin{equation}
  \mathcal L_{\mathrm{CTR}}=-\frac1{|\mathcal B|}\sum_{b\in\mathcal B}
       \bigl[y_b\log\hat y_b+(1-y_b)\log(1-\hat y_b)\bigr].
  \label{eq:trip-ctr-loss}
\end{equation}
The reconstruction objective encourages contextualized event representations to recover masked item content:
\begin{equation}
  \mathcal L_{\mathrm{rec}}=\frac1{|\mathcal M_{\mathcal B}|}
    \sum_{i\in\mathcal M_{\mathcal B}}
    \left[1-\cos\left(\widehat{\mathbf e}_i,
               \operatorname{sg}(\mathbf e_i^{\mathrm{item}})\right)\right],
  \label{eq:trip-rec-loss}
\end{equation}
where $\mathcal M_{\mathcal B}$ denotes the masked events in the minibatch, $\widehat{\mathbf e}_i$ is the reconstructed item embedding, and $\operatorname{sg}$ stops gradients through the original embedding used as the target. The joint objective is
\begin{equation}
  \mathcal L=\mathcal L_{\mathrm{CTR}}+\lambda_{\mathrm{rec}}\mathcal L_{\mathrm{rec}},
  \label{eq:trip-joint-loss}
\end{equation}
where $\lambda_{\mathrm{rec}}$ controls the auxiliary supervision strength. The model parameters are jointly optimized under this objective. Training and implementation details are provided in Appendix~\ref{app:implementation}.

\stitle{Limitations.}
TRIP defines temporal groups using fixed logarithmic intervals relative to the prediction time, which may not align with individual users' interest transitions. In addition, mean-pooling contextualized events into group summaries can obscure fine-grained event distinctions in subsequent layers. Adaptive group boundaries and selective retention of event-level representations could address these limitations.

\begin{table}[t]
\centering
\caption{Dataset statistics and input feature configurations.}
\label{tab:dataset-statistics}
\setlength{\tabcolsep}{3pt}
\renewcommand{\arraystretch}{1.1}
\small
\resizebox{0.95\linewidth}{!}{\begin{tabular}{@{}lrrrr@{}}
\toprule
Statistic & KuaiRec & KuaiRand & TAAC & TaobaoAd \\
\midrule
Train samples & 12,190,187 & 10,911,770 & 4,714,200 & 20,015,245 \\
Val. samples & 181,130 & 390,961 & 148,699 & 3,234,051 \\
Test samples & 159,040 & 410,314 & 92,495 & 3,308,665 \\
\midrule
User ID vocab. & 7,176 & 1,000 & 1,001,845 & 1,141,729 \\
Item ID vocab. & 10,728 & 4,369,953 & 4,783,154 & 12,975 \\
\midrule
User fields & 31 & 31 & 9 & 9 \\
Target fields & 45 & 50 & 14 & 5 \\
\midrule
Max. history & 200 & 200 & 100 & 300 \\
Train pos. rate & 7.44\% & 37.83\% & 10.92\% & 5.20\% \\
\bottomrule
\end{tabular}}
\end{table}

\begin{table*}[t]
\centering
\caption{Overall performance comparison on the public KuaiRec, KuaiRand, TAAC, and TaobaoAd datasets in terms of AUC and LogLoss. Best results are bolded and second-best results are underlined.}
\label{tab:main_results}
\newcommand{\mainstd}[1]{\,\mbox{\scriptsize$\pm #1$}}
\normalsize
\renewcommand{\arraystretch}{1.1}
\resizebox{0.95\textwidth}{!}{\begin{tabular}{l*{8}{c}}
\toprule
 & \multicolumn{2}{c}{KuaiRec} & \multicolumn{2}{c}{KuaiRand} & \multicolumn{2}{c}{TAAC} & \multicolumn{2}{c}{TaobaoAd} \\
\cmidrule(lr){2-3}\cmidrule(lr){4-5}\cmidrule(lr){6-7}\cmidrule(lr){8-9}
Model & AUC $\uparrow$ & LogLoss $\downarrow$ & AUC $\uparrow$ & LogLoss $\downarrow$ & AUC $\uparrow$ & LogLoss $\downarrow$ & AUC $\uparrow$ & LogLoss $\downarrow$ \\
\midrule
DeepFM & $0.8329\mainstd{0.0021}$ & $0.2261\mainstd{0.0015}$ & $0.7362\mainstd{0.0016}$ & $0.5838\mainstd{0.0008}$ & $0.5729\mainstd{0.0047}$ & $0.1914\mainstd{0.0030}$ & $0.5902\mainstd{0.0024}$ & $0.2119\mainstd{0.0009}$ \\
DCNv2 & $0.8445\mainstd{0.0016}$ & $0.2262\mainstd{0.0185}$ & $0.7730\mainstd{0.0012}$ & $\underline{0.5487}\mainstd{0.0018}$ & $0.6018\mainstd{0.0016}$ & $0.1838\mainstd{0.0042}$ & $0.6232\mainstd{0.0013}$ & $0.2173\mainstd{0.0086}$ \\
WuKong & $0.8448\mainstd{0.0012}$ & $0.2191\mainstd{0.0015}$ & $0.7739\mainstd{0.0028}$ & $0.5506\mainstd{0.0040}$ & $0.6021\mainstd{0.0027}$ & $0.1820\mainstd{0.0054}$ & $0.6251\mainstd{0.0003}$ & $0.1974\mainstd{0.0012}$ \\
RankMixer & $0.8449\mainstd{0.0001}$ & $0.2197\mainstd{0.0031}$ & $0.7752\mainstd{0.0008}$ & $\mathbf{0.5481}\mainstd{0.0002}$ & $0.6021\mainstd{0.0021}$ & $0.1862\mainstd{0.0043}$ & $0.6241\mainstd{0.0003}$ & $0.1983\mainstd{0.0013}$ \\
\midrule
DIN & $0.8447\mainstd{0.0009}$ & $0.2154\mainstd{0.0004}$ & $0.7705\mainstd{0.0003}$ & $0.5518\mainstd{0.0010}$ & $0.6042\mainstd{0.0019}$ & $0.1875\mainstd{0.0052}$ & $0.6235\mainstd{0.0008}$ & $0.1977\mainstd{0.0011}$ \\
DIEN & $0.8449\mainstd{0.0015}$ & $0.2142\mainstd{0.0011}$ & $0.7686\mainstd{0.0007}$ & $0.5593\mainstd{0.0032}$ & $0.6018\mainstd{0.0011}$ & $0.1795\mainstd{0.0029}$ & $0.6257\mainstd{0.0009}$ & $0.1965\mainstd{0.0011}$ \\
STCA & $0.8440\mainstd{0.0023}$ & $0.2161\mainstd{0.0004}$ & $0.7705\mainstd{0.0005}$ & $0.5629\mainstd{0.0106}$ & $0.6014\mainstd{0.0008}$ & $0.1888\mainstd{0.0026}$ & $0.6228\mainstd{0.0005}$ & $0.1974\mainstd{0.0006}$ \\
\midrule
TokenFormer & $0.8437\mainstd{0.0011}$ & $0.2181\mainstd{0.0007}$ & $0.7593\mainstd{0.0013}$ & $0.5794\mainstd{0.0062}$ & $0.5981\mainstd{0.0060}$ & $0.1838\mainstd{0.0046}$ & $0.6263\mainstd{0.0005}$ & $0.1964\mainstd{0.0007}$ \\
OneTrans & $0.8443\mainstd{0.0012}$ & $0.2163\mainstd{0.0005}$ & $0.7694\mainstd{0.0013}$ & $0.5554\mainstd{0.0020}$ & $0.6067\mainstd{0.0011}$ & $0.1857\mainstd{0.0061}$ & $0.6276\mainstd{0.0015}$ & $0.1966\mainstd{0.0002}$ \\
HyFormer & $0.8449\mainstd{0.0005}$ & $0.2151\mainstd{0.0006}$ & $0.7683\mainstd{0.0016}$ & $0.5559\mainstd{0.0008}$ & $0.6050\mainstd{0.0009}$ & $0.1826\mainstd{0.0023}$ & $0.6254\mainstd{0.0012}$ & $\underline{0.1962}\mainstd{0.0005}$ \\
MixFormer & $0.8438\mainstd{0.0034}$ & $0.2210\mainstd{0.0058}$ & $\underline{0.7754}\mainstd{0.0006}$ & $0.5490\mainstd{0.0014}$ & $0.6042\mainstd{0.0012}$ & $\underline{0.1793}\mainstd{0.0038}$ & $0.6265\mainstd{0.0008}$ & $0.1977\mainstd{0.0004}$ \\
\midrule
\rowcolor{black!5}
\textbf{TRIP-B} & $\mathbf{0.8512}\mainstd{0.0016}$ & $\mathbf{0.2114}\mainstd{0.0007}$ & $\mathbf{0.7778}\mainstd{0.0008}$ & $0.5692\mainstd{0.0076}$ & $\mathbf{0.6337}\mainstd{0.0017}$ & $0.1802\mainstd{0.0084}$ & $\mathbf{0.6299}\mainstd{0.0007}$ & $0.1964\mainstd{0.0002}$ \\
\rowcolor{black!5}
\textbf{TRIP-R} & $\underline{0.8506}\mainstd{0.0012}$ & $\underline{0.2116}\mainstd{0.0011}$ & $0.7731\mainstd{0.0008}$ & $0.5653\mainstd{0.0024}$ & $\underline{0.6331}\mainstd{0.0013}$ & $\mathbf{0.1753}\mainstd{0.0087}$ & $\underline{0.6292}\mainstd{0.0014}$ & $\mathbf{0.1958}\mainstd{0.0009}$ \\
\bottomrule
\end{tabular}}
\end{table*}

\section{Experiments}
\label{sec:experiments}

\subsection{Experimental Setup}
\label{sec:experimental-setup}

\stitle{Datasets.}
We evaluate on four public benchmarks. KuaiRec~\cite{gao2022kuairec} contains short-video viewing feedback with a compact catalog; KuaiRand-1K~\cite{gao2022kuairand} provides short-video logs with randomized exposure interventions and a larger catalog. TAAC-1M~\cite{pan2026taac2025} contains Tencent advertising sequences, while TaobaoAd~\cite{alimamaTaobaoAd} combines display-ad impressions with shopping histories. These settings cover different recommendation domains, catalog sizes, history lengths, and feedback distributions (Table~\ref{tab:dataset-statistics}).

\stitle{Baselines.}
We use two sets of baselines to evaluate overall model performance and the contribution of positional encoding.

\begin{itemize}[leftmargin=*,topsep=2pt,itemsep=3pt,parsep=0pt]
  \item \textbf{Overall model performance.} We compare TRIP with eleven models: (1) feature-interaction models, including DeepFM~\cite{guo2017deepfm}, DCNv2~\cite{wang2021dcnv2}, Wukong~\cite{zhang2024wukong}, and RankMixer~\cite{zhu2025rankmixer}; (2) behavior-sequence models, including DIN~\cite{zhou2018din}, DIEN~\cite{zhou2019dien}, and STCA~\cite{guan2026stca}; and (3) unified feature and sequence models, including TokenFormer~\cite{zhou2026tokenformer}, OneTrans~\cite{zhang2026onetrans}, HyFormer~\cite{huang2026hyformer}, and MixFormer~\cite{huang2026mixformer}.
  \item \textbf{Positional encoding.} We compare FRE + TRTE-B and FRE + TRTE-R with eight configurations in four categories: (1) physical-position configurations, including the NoPE reference, learned APE, sinusoidal APE, and rank RoPE~\cite{su2024roformer}; (2) time-based encodings, including T-RoPE~\cite{liu2026trope} and ClockRoPE~\cite{chen2026clockrope}; (3) target-aware encoding, represented by CAPE~\cite{yuan2025cape}; and (4) joint time--position encoding, represented by TO-RoPE~\cite{wei2026torope}. We evaluate them on OneTrans, HyFormer, and MixFormer to assess the contribution and transferability of FRE and TRTE independently of TGCI.
\end{itemize}

\stitle{Training and Evaluation.}
We implement all models in PyTorch and use the same NVIDIA H20 hardware, optimizer, learning rate, batch sizes, and training duration for fair comparison. All baselines retain raw event-time and prediction-time features, so the comparisons do not withhold timestamp information from existing methods. We evaluate predictive performance using AUC and LogLoss, reporting the mean and sample standard deviation over three random seeds. Further training, model, and evaluation details are provided in Appendix~\ref{app:training-details}.

\subsection{Overall Performance}
\label{sec:overall-performance}

Table~\ref{tab:main_results} shows that TRIP-B consistently outperforms the eleven baselines in AUC across all four datasets, with an average relative improvement of 1.47\% and a maximum of 4.45\% on TAAC. TRIP also achieves the lowest LogLoss on KuaiRec, TAAC, and TaobaoAd, demonstrating strong ranking discrimination and probabilistic prediction performance. These gains reflect the complementary roles of FRE and TRTE in modeling behavioral rank and target-relative time, while TGCI enriches behaviors within target-relative temporal groups and preserves interest evolution across groups. Together, these components provide richer positional context and more effective historical interaction. On KuaiRand, despite its AUC advantage, TRIP-B exhibits a higher LogLoss. KuaiRand contains heterogeneous exposure scenarios with distinct feedback definitions and positive rates, and analysis reveals that TRIP-B substantially overestimates positive-feedback probabilities in certain low-positive-rate scenarios, accounting for much of the LogLoss gap (Appendix~\ref{app:kuairand-calibration}).

\begin{table*}[t]
\centering
\caption{Positional encoding transfer across three CTR backbones (mean AUC). FRE + TRTE is evaluated without TGCI. Best and second-best results are bolded and underlined.}
\label{tab:encoding-transfer}
\small
\setlength{\tabcolsep}{1.8pt}
\renewcommand{\arraystretch}{1.15}

\resizebox{0.95\textwidth}{!}{\begin{tabular}{@{}l*{12}{c}@{}}
\toprule
\multirow{2}{*}{\shortstack[l]{Positional\\Encoding}} & \multicolumn{3}{c}{KuaiRec} & \multicolumn{3}{c}{KuaiRand} & \multicolumn{3}{c}{TAAC} & \multicolumn{3}{c}{TaobaoAd} \\
\cmidrule(lr){2-4}\cmidrule(lr){5-7}\cmidrule(lr){8-10}\cmidrule(lr){11-13}
 & {\small OneTrans} & {\small HyFormer} & {\small MixFormer} & {\small OneTrans} & {\small HyFormer} & {\small MixFormer} & {\small OneTrans} & {\small HyFormer} & {\small MixFormer} & {\small OneTrans} & {\small HyFormer} & {\small MixFormer} \\
\midrule
NoPE & $0.8443$ & $0.8449$ & $0.8438$ & $0.7694$ & $0.7683$ & $\underline{0.7754}$ & $0.6067$ & $0.6050$ & $0.6042$ & $0.6276$ & $0.6254$ & $0.6265$ \\
Learned APE & $0.8445$ & $0.8466$ & $0.8439$ & $0.7589$ & $0.7621$ & $0.7660$ & $0.6058$ & $0.6037$ & $0.6034$ & $0.6281$ & $0.6278$ & $0.6275$ \\
Sinusoidal APE & $0.8457$ & $0.8450$ & $0.8456$ & $0.7648$ & $0.7641$ & $0.7645$ & $0.6061$ & $0.6038$ & $0.6063$ & $0.6273$ & $0.6276$ & $0.6279$ \\
Rank RoPE & $0.8434$ & $0.8432$ & $0.8442$ & $0.7616$ & $0.7623$ & $0.7026$ & $0.6057$ & $0.6018$ & $0.6058$ & $0.6280$ & $0.6267$ & $0.6273$ \\
CAPE & $0.8424$ & $0.8421$ & $0.8189$ & $0.7641$ & $0.7641$ & $0.6418$ & $0.6055$ & $0.6059$ & $0.5729$ & $0.6266$ & $0.6248$ & $0.6106$ \\
TO-RoPE & $0.8474$ & $0.8462$ & $0.8471$ & $0.7668$ & $0.7640$ & $0.6742$ & $0.6062$ & $0.6031$ & $0.6061$ & $0.6283$ & $0.6260$ & $0.6269$ \\
T-RoPE & $0.8497$ & $0.8453$ & $0.8452$ & $0.7668$ & $0.7687$ & $0.7675$ & $0.6054$ & $0.6035$ & $0.6021$ & $\underline{0.6292}$ & $0.6261$ & $0.6225$ \\
ClockRoPE & $0.8488$ & $0.8464$ & $\underline{0.8496}$ & $0.7658$ & $0.7643$ & $0.7648$ & $0.6056$ & $0.6036$ & $0.6053$ & $0.6284$ & $0.6274$ & $0.6267$ \\
\midrule
\rowcolor{black!5}
FRE + TRTE-B & $\mathbf{0.8504}$ & $\mathbf{0.8502}$ & $0.8495$ & $\mathbf{0.7756}$ & $\underline{0.7693}$ & $\mathbf{0.7758}$ & $\underline{0.6319}$ & $\underline{0.6137}$ & $\underline{0.6277}$ & $0.6291$ & $\underline{0.6279}$ & $\mathbf{0.6286}$ \\
\rowcolor{black!5}
FRE + TRTE-R & $\underline{0.8502}$ & $\underline{0.8490}$ & $\mathbf{0.8528}$ & $\underline{0.7747}$ & $\mathbf{0.7708}$ & $0.7719$ & $\mathbf{0.6322}$ & $\mathbf{0.6288}$ & $\mathbf{0.6284}$ & $\mathbf{0.6296}$ & $\mathbf{0.6285}$ & $\underline{0.6282}$ \\
\bottomrule
\end{tabular}}
\end{table*}

\subsection{Evaluation of Positional Encoding}
\label{sec:encoding-evaluation}

Table~\ref{tab:encoding-transfer} evaluates FRE + TRTE against eight positional and temporal configurations on three CTR backbones, without using TGCI. The additive and rotary variants outperform the strongest competing encoding in 10 and 11 of the 12 settings, respectively, with at least one achieving the highest AUC in every setting. On TAAC, FRE + TRTE-R improves AUC by 4.20\%, 3.78\%, and 3.65\% on OneTrans, HyFormer, and MixFormer, respectively. These results demonstrate that the proposed rank--time representation is transferable across backbones independently of the complete TRIP pipeline.
The comparisons further highlight the limitations of existing positional encodings. On KuaiRand, learned APE, sinusoidal APE, and rank RoPE underperform NoPE across all three backbones, while time-based T-RoPE and ClockRoPE also trail NoPE on OneTrans. On TAAC, target-aware CAPE and joint order--time TO-RoPE remain below FRE + TRTE across all three backbones. These findings suggest that incorporating positional cues does not by itself guarantee effective modeling. FRE + TRTE instead couples task-adaptive rank representations with prediction-time-anchored temporal encoding, jointly interpreting sequence rank and event time relative to the current prediction.

\raggedbottom
\stitle{Rank Encoding Replacements.}
Table~\ref{tab:rank-encoding-replacements} replaces FRE in TRIP-B with other rank encodings while keeping TRTE fixed. The contrast is sharpest on TAAC. Rank encoding alone does not improve over NoPE there, and removing FRE from TRIP-B returns AUC to a similar level, yet with TRTE in place, learned APE and FRE outperform Rank RoPE by about 0.027. Rank RoPE still exposes recency, because histories are right-aligned and prediction-side tokens share a fixed rotary coordinate, but it only changes attention scores. Additive encodings instead write rank into the token representations that TRTE aggregates by target-relative time, consistent with rank and time being complementary when both enter the representations. FRE performs comparably to or better than full learned APE on all four datasets while using 3.8--21.6$\times$ fewer positional parameters, since each rank is expressed through a few shared basis vectors.

\begin{table}[!htbp]
  \centering
  \caption{Rank encoding replacements in TRIP-B with TRTE fixed. The best mean in each column is bolded.}
  \label{tab:rank-encoding-replacements}
  \footnotesize
  \setlength{\tabcolsep}{2pt}
  \renewcommand{\arraystretch}{1.1}
  \newcommand{\rankcell}[2]{$#1${\fontsize{6.5}{7}\selectfont$\,\pm\,#2$}}
  \resizebox{0.95\linewidth}{!}{\begin{tabular*}{\columnwidth}{@{\extracolsep{\fill}}l*{4}{c}@{}}
    \toprule
    Encoding & KuaiRec & KuaiRand & TAAC & TaobaoAd \\
    \midrule
    FRE & \rankcell{\mathbf{0.8512}}{\mathbf{0.0016}} & \rankcell{0.7778}{0.0008} & \rankcell{\mathbf{0.6337}}{\mathbf{0.0017}} & \rankcell{\mathbf{0.6299}}{\mathbf{0.0007}} \\
    Learned APE & \rankcell{0.8493}{0.0009} & \rankcell{\mathbf{0.7787}}{\mathbf{0.0012}} & \rankcell{0.6323}{0.0015} & \rankcell{0.6297}{0.0013} \\
    Sinusoidal APE & \rankcell{0.8504}{0.0010} & \rankcell{0.7782}{0.0009} & \rankcell{0.6240}{0.0121} & \rankcell{0.6292}{0.0012} \\
    Rank RoPE & \rankcell{0.8496}{0.0014} & \rankcell{0.7753}{0.0010} & \rankcell{0.6061}{0.0021} & \rankcell{0.6287}{0.0017} \\
    \bottomrule
  \end{tabular*}}
\end{table}

\subsection{Ablation and Mechanism Analysis}
\label{sec:ablation-study}

\begin{figure}[!htbp]
  \centering
  \includegraphics[width=0.95\linewidth]{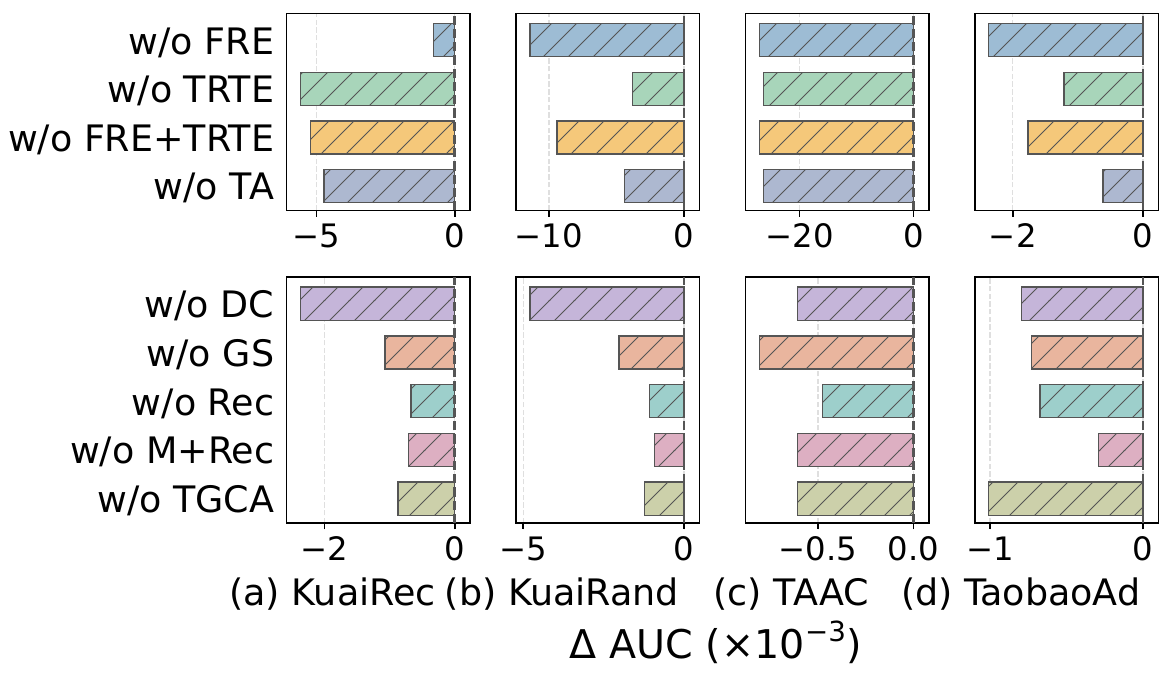}
  \caption{Ablations of TRIP-B. Bars show mean AUC differences from Full ($\times10^3$). TA, DC, GS, Rec, and M denote target awareness, density correction, group summaries, reconstruction, and masking; TGCA denotes attention within TGCI.}
  \Description{Core ablations on four datasets examine rank, temporal encoding, target awareness, density correction, summaries, reconstruction, masking, and group-causal interaction through mean AUC differences from full TRIP-B.}
  \label{fig:core-ablation}
\end{figure}

Figure~\ref{fig:core-ablation} evaluates nine ablated variants of TRIP-B. The full model achieves the highest mean AUC among the displayed variants on all four datasets, demonstrating the benefits of combining rank--time positional modeling with temporal-group interaction.

\stitle{Ordinal and Target-Relative Information.}
Removing either FRE or TRTE lowers AUC across all four datasets, indicating that each contributes beyond the other. The effect is particularly pronounced on TAAC, where AUC drops from 0.6337 to 0.6066 without FRE and 0.6073 without TRTE. Replacing the prediction-time reference while keeping the other components fixed also reduces AUC across datasets. These results support retaining both ordinal and temporal information, with the current prediction timestamp providing an effective reference for temporal modeling.

\begin{figure*}[t]
  \centering
  \includegraphics[width=0.95\textwidth]{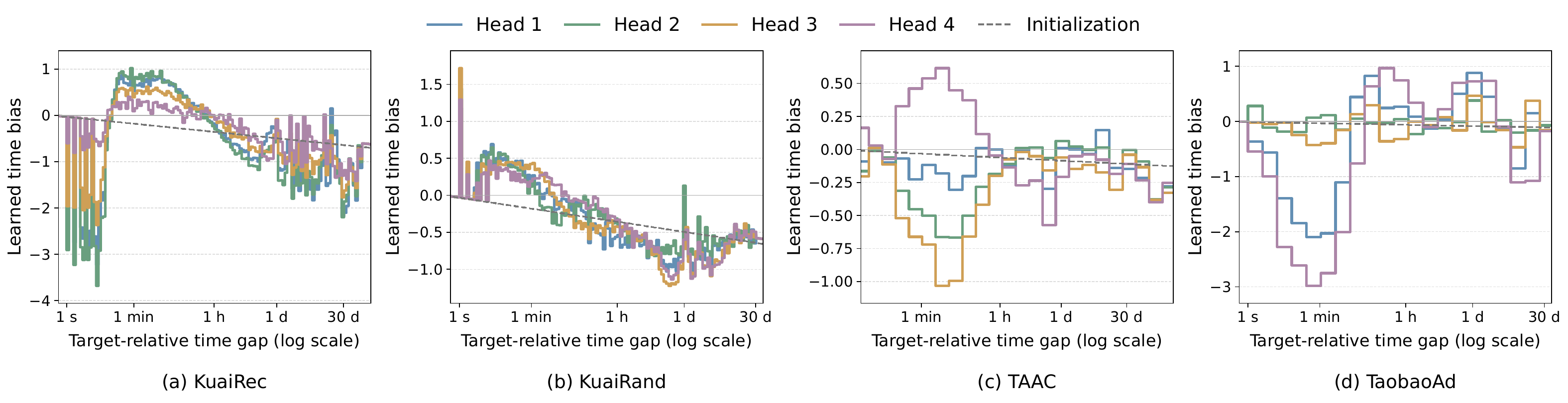}
  \caption{Learned temporal potentials $\phi_h$ on four datasets. Curves denote attention heads; dashed lines show the original initialization. Horizontal axes show target-relative elapsed time on a logarithmic scale.}
  \Description{Four panels show head-specific learned temporal preferences on KuaiRec, KuaiRand, TAAC, and TaobaoAd before density correction.}
  \label{fig:learned-temporal-potentials}
\end{figure*}

\stitle{Temporal Preferences and Density.}
Removing density correction lowers mean AUC across all four datasets, supporting occupancy adjustment to counteract the attention advantage of densely populated intervals. Figure~\ref{fig:learned-temporal-potentials} further shows dataset- and head-specific temporal patterns: the learned potentials assign higher scores to minute-scale intervals on KuaiRec, exhibit a broad recency preference on KuaiRand, and differ across heads on TAAC. TaobaoAd shows separated high-potential windows rather than uniform recency decay. These curves describe temporal biases, not final attention weights, which also depend on content scores and density correction. Together, these observations support flexible temporal preferences with sample-wise occupancy adjustment.

\stitle{Within-Group Interaction and Group-Level Modeling.}
Replacing group-causal interaction with token-level causal attention lowers mean AUC on every dataset even when temporal grouping and summaries are retained, supporting the contribution of within-group bidirectional context. Removing summaries also reduces performance, indicating the value of group-level aggregation after event contextualization. Removing reconstruction while retaining masking further lowers AUC, supporting the contribution of auxiliary reconstruction supervision. These controlled comparisons support the roles of within-group enrichment, group summaries, and reconstruction within TGCI. Variant definitions and parameter sensitivity results are given in Appendices~\ref{app:ablation-details} and~\ref{app:parameter-sensitivity}.

\begin{figure}[t]
  \centering
  \includegraphics[width=0.95\linewidth]{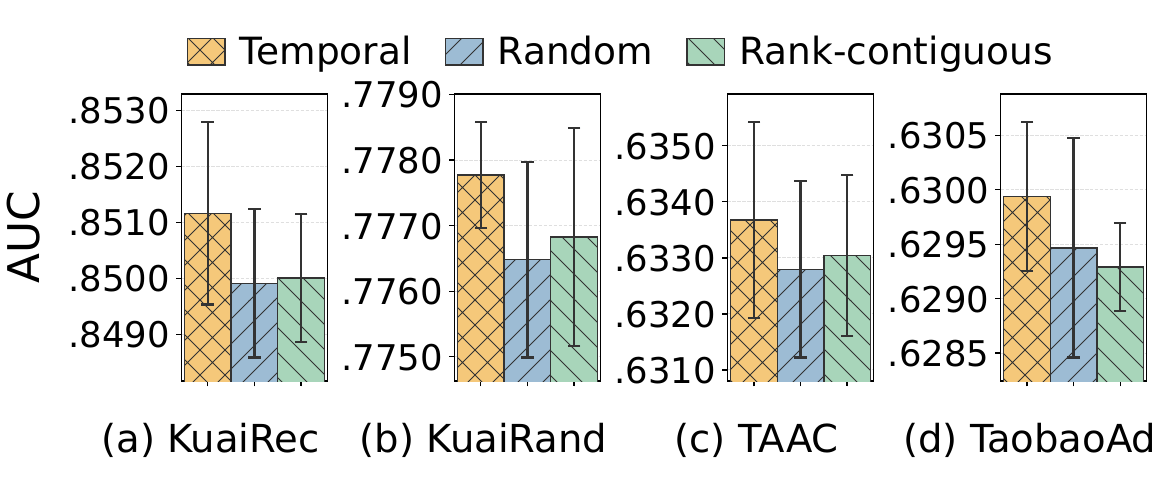}
  \caption{Mean AUC under temporal, random, and rank-contiguous grouping. All strategies preserve the original group counts and sizes for each sample.}
  \Description{Four panels compare the grouping strategies on KuaiRec, KuaiRand, TAAC, and TaobaoAd. Temporal grouping has the highest mean AUC in each panel. Bars include error bars, and the panels use different AUC scales.}
  \label{fig:grouping-comparison}
\end{figure}

\stitle{Grouping Strategies.}
Figure~\ref{fig:grouping-comparison} compares temporal grouping with Random and Rank-contiguous alternatives. All strategies preserve per-sample group counts and sizes, keeping the summary compression ratio unchanged. Random redistributes group membership, whereas Rank-contiguous partitions consecutive ranks using reordered temporal-group sizes. Under these matched conditions, temporal grouping achieves the highest mean AUC on all four datasets. This suggests that grouping behaviors by their distance to the prediction is more effective than arbitrary or rank-based partitions, beyond the benefit of compression alone. Construction details are given in Appendix~\ref{app:grouping-details}.

\subsection{Efficiency of Group-Level Modeling}
\label{sec:summary-analysis}

Table~\ref{tab:summary-efficiency} shows two benefits of group-summary compression: higher AUC and 6.3--32.8\% lower per-step training time. On TaobaoAd, AUC improves from 0.6297 to 0.6304 while time falls from 141.31 to 95.01 ms/step; KuaiRand gains AUC from 0.7753 to 0.7778 with 20.7\% time savings. The AUC gains suggest that aggregation suppresses redundant or noisy event-level variation while retaining useful interest signals. Preserving event-level inputs in the first Transformer block allows the model to learn contextual representations from the original behavior features before compression. Subsequent aggregation can therefore consolidate the contextualized interest of each group rather than discard event details before they are encoded. The shorter group-level sequence also reduces later-layer computation, improving accuracy and efficiency together. Further timing comparisons are provided in Appendix~\ref{app:efficiency-details}.

\begin{table}[!htbp]
  \centering
  \caption{Analysis of Group-summary efficiency (single-run results). Time savings are relative to no compression.}
  \label{tab:summary-efficiency}
  \small
  \setlength{\tabcolsep}{4pt}
  \renewcommand{\arraystretch}{1.12}
  \begin{tabular}{@{}lccc@{}}
    \toprule
    Dataset & AUC (w/o $\rightarrow$ with) & Saved & ms/step \\
    \midrule
    KuaiRec & $0.8501\rightarrow0.8508$ & $20.0\%$ & $113.95$ \\
    KuaiRand & $0.7753\rightarrow0.7778$ & $20.7\%$ & $116.54$ \\
    TAAC & $0.6355\rightarrow0.6359$ & $6.3\%$ & $102.13$ \\
    TaobaoAd & $0.6297\rightarrow0.6304$ & $32.8\%$ & $95.01$ \\
    \bottomrule
  \end{tabular}
\end{table}

\section{Related Work}
\label{sec:related-work}
\stitle{CTR Modeling.}
CTR modeling has developed along two lines: feature interaction and sequence modeling. Feature models have progressed from factorization-based and explicit crosses~\cite{guo2017deepfm,wang2021dcnv2} to attention-based interactions~\cite{song2019autoint}. Wukong~\cite{zhang2024wukong} and RankMixer~\cite{zhu2025rankmixer} scale interactions through stacked factorization and token mixing, respectively. Sequence models have evolved from target-conditioned interest aggregation and evolution~\cite{zhou2018din,zhou2019dien} toward efficient long-history processing with LONGER~\cite{chai2025longer} and STCA~\cite{guan2026stca}. These separate modules increasingly converge toward unified architectures for joint optimization and sequence--feature interaction~\cite{zhang2026onetrans,huang2026hyformer,huang2026mixformer,zhou2026tokenformer,liu2026est}. OneTrans~\cite{zhang2026onetrans} processes historical and non-sequential tokens with causal Transformer blocks. HyFormer~\cite{huang2026hyformer} and MixFormer~\cite{huang2026mixformer} further explore the integration and joint scaling of sequence modeling and feature interaction. However, suitable positional encoding and interaction granularity for Transformer-based Unified CTR prediction remain unclear.

\stitle{Position Encoding for Sequential Recommendation.}
Positional encodings represent sequence indices, event time, target relevance, or joint order--time relationships~\cite{su2024roformer,li2020tisasrec,yuan2025cape,wei2026torope}. Index-based learned or rotary encodings remain common~\cite{kang2018sasrec,sun2019bert4rec,su2024roformer,zhang2026onetrans,zhou2026tokenformer}. Other methods incorporate event timing or intervals to account for irregular behavior histories~\cite{li2020tisasrec,liu2026trope,chen2026clockrope,wei2026torope,pmlr-v235-zhai24a}. For example, T-RoPE~\cite{liu2026trope} uses multiscale timestamp rotations with prediction-time query alignment, while TO-RoPE~\cite{wei2026torope} combines index and time rotations within shared or separate rotary subspaces. In addition, context-aware CAPE~\cite{yuan2025cape} constructs soft positions from target--history item dissimilarity, expressing item relevance rather than elapsed time. TRIP combines task-adaptive rank representations with prediction-time-anchored temporal encoding and target-relative temporal grouping.

\section{Conclusion}
In this work, we found that existing positional encodings do not consistently improve CTR prediction, while bidirectional interaction within target-relative temporal groups can outperform both token-level causal and global bidirectional attention. These observations reveal the need to reconsider both historical position representations and interaction granularity. Based on these findings, we proposed TRIP, which jointly models sequence rank, event time, and the prediction-time reference, and organizes history into target-relative temporal groups with intra-group bidirectional and inter-group causal interaction. Experiments on four public datasets demonstrate TRIP's effectiveness, while positional-encoding transfer experiments across three unified backbones validate the portability of its rank--time design. Detailed ablations and representation analyses further support the component designs.

\flushbottom
\bibliographystyle{ACM-Reference-Format}
\bibliography{sample-base}

\appendix

\section{Implementation Details}
\label{app:implementation}

\subsection{Training and Evaluation Settings}
\label{app:training-details}
\stitle{Common Training Settings.}
We implement all models in PyTorch and train each run on one NVIDIA H20 GPU. TRIP and all baselines share AdamW, a fixed learning rate of $10^{-3}$, weight decay $10^{-2}$, training/evaluation batch sizes of 1024/2048, and 64-dimensional sparse feature embeddings. All models are trained for one epoch with BF16 mixed precision and gradient-norm clipping at 1.0. This single-pass protocol limits the risk of overfitting sparse embeddings through repeated exposure to the same impressions. Hyperparameters are selected using only training and validation data, and checkpoints are selected by validation AUC before final test evaluation. We report the mean and sample standard deviation over seeds 41, 42, and 43, counting only real samples. These shared training and evaluation settings ensure a consistent comparison. The TRIP implementation uses PyTorch 2.5.1, CUDA 12.1, and cuBLAS 12.8.5.5; timing protocols are detailed in Appendix~\ref{app:efficiency-details}.

\stitle{Model and Hyperparameter Settings.}
\label{app:hyperparameters}
TRIP uses four Transformer blocks, hidden dimension 64, four attention heads, and FFN dimension 256. The prediction head mean-pools the final prediction-side states and uses a two-layer MLP. The non-sequential token counts $N_{\mathrm{ns}}$ are 12, 18, 12, and 12 for KuaiRec, KuaiRand, TAAC, and TaobaoAd, respectively. Most Transformer-based baselines share these block, hidden, attention-head, and FFN settings while retaining their respective architectures. RankMixer~\cite{zhu2025rankmixer} retains its 16-head configuration, while MixFormer~\cite{huang2026mixformer} uses a per-head dimension of 64. Method-specific hyperparameters follow the original model designs and are tuned on validation data. Table~\ref{tab:trip-hyperparameters} lists TRIP's selected settings. FRE and reconstruction settings are shared by both variants, whereas $m_{\mathrm B}$, $J_{\max}$, and $\lambda_{\mathrm{cnt}}$ apply to TRIP-B and $m_{\mathrm R}$ to TRIP-R. Temporal grouping uses $C_{\mathrm{grp}}=32$ independently of encoding resolution, and TRIP-R uses $C_{\mathrm{rot}}=32$.

\stitle{Data and Sequence Processing.}
Histories retain positive-feedback events on KuaiRec and KuaiRand and all events on TAAC and TaobaoAd. KuaiRec uses watch ratio $\ge2$ for positive feedback; TaobaoAd uses category IDs for history--target matching, with adgroup ID as a separate feature. For each target impression, valid history events are ordered chronologically and truncated to the most recent $L$ events. Shorter histories are left-padded, with valid events right-aligned: the event nearest the prediction-side tokens is the most recent, and events farther to the left occurred earlier. Padding is excluded by the validity mask. All models use the same dataset-specific history construction.

\begin{table}[t]
  \centering
  \caption{Dataset-specific FRE, reconstruction, and time-encoding settings.}
  \label{tab:trip-hyperparameters}
  \small
  \setlength{\tabcolsep}{2pt}
  \renewcommand{\arraystretch}{1.15}
  \resizebox{0.85\linewidth}{!}{\begin{tabular}{@{}l*{9}{c}@{}}
    \toprule
    Dataset & $K$ & $\gamma_\mu$ & $\gamma_R$ & $\rho$ & $\lambda_{\mathrm{rec}}$ & $m_{\mathrm B}$ & $J_{\max}$ & $\lambda_{\mathrm{cnt}}$ & $m_{\mathrm R}$ \\
    \midrule
    KuaiRec & 4 & $1.0$ & $1.0$ & $0.15$ & $10^{-4}$ & 6 & 192 & $0.50$ & 6 \\
    KuaiRand & 2 & $0.75$ & $1.0$ & $0.05$ & $0.01$ & 8 & 256 & $0.75$ & 6 \\
    TAAC & 10 & $0.25$ & $1.0$ & $0.15$ & $0.01$ & 1 & 32 & $0.50$ & 2 \\
    TaobaoAd & 4 & $1.0$ & $1.0$ & $0.05$ & $0.01$ & 1 & 32 & $0.50$ & 1 \\
    \bottomrule
  \end{tabular}}
\end{table}

\begin{figure}[t]
  \centering
  \includegraphics[width=0.49\linewidth]{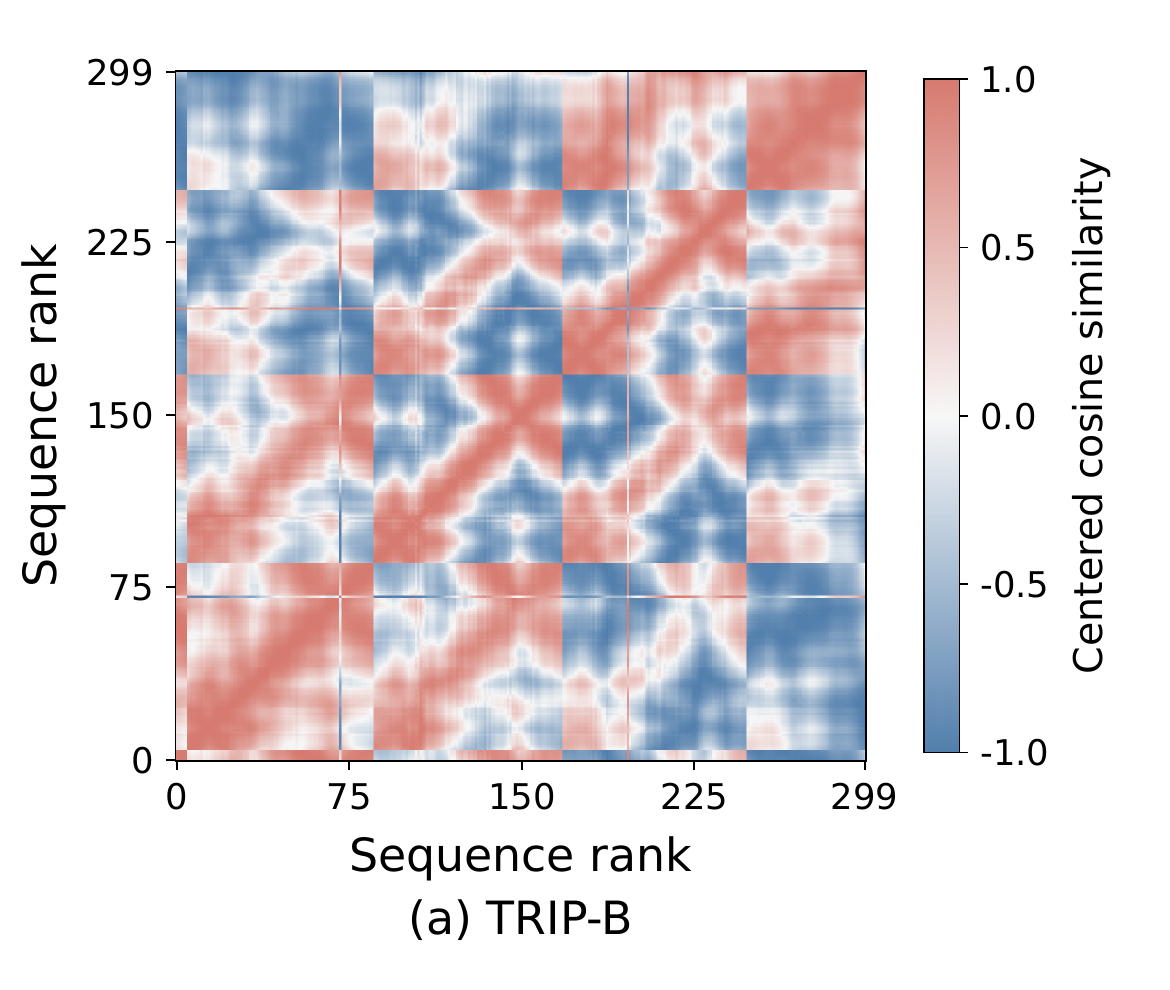}
  \hfill
  \includegraphics[width=0.49\linewidth]{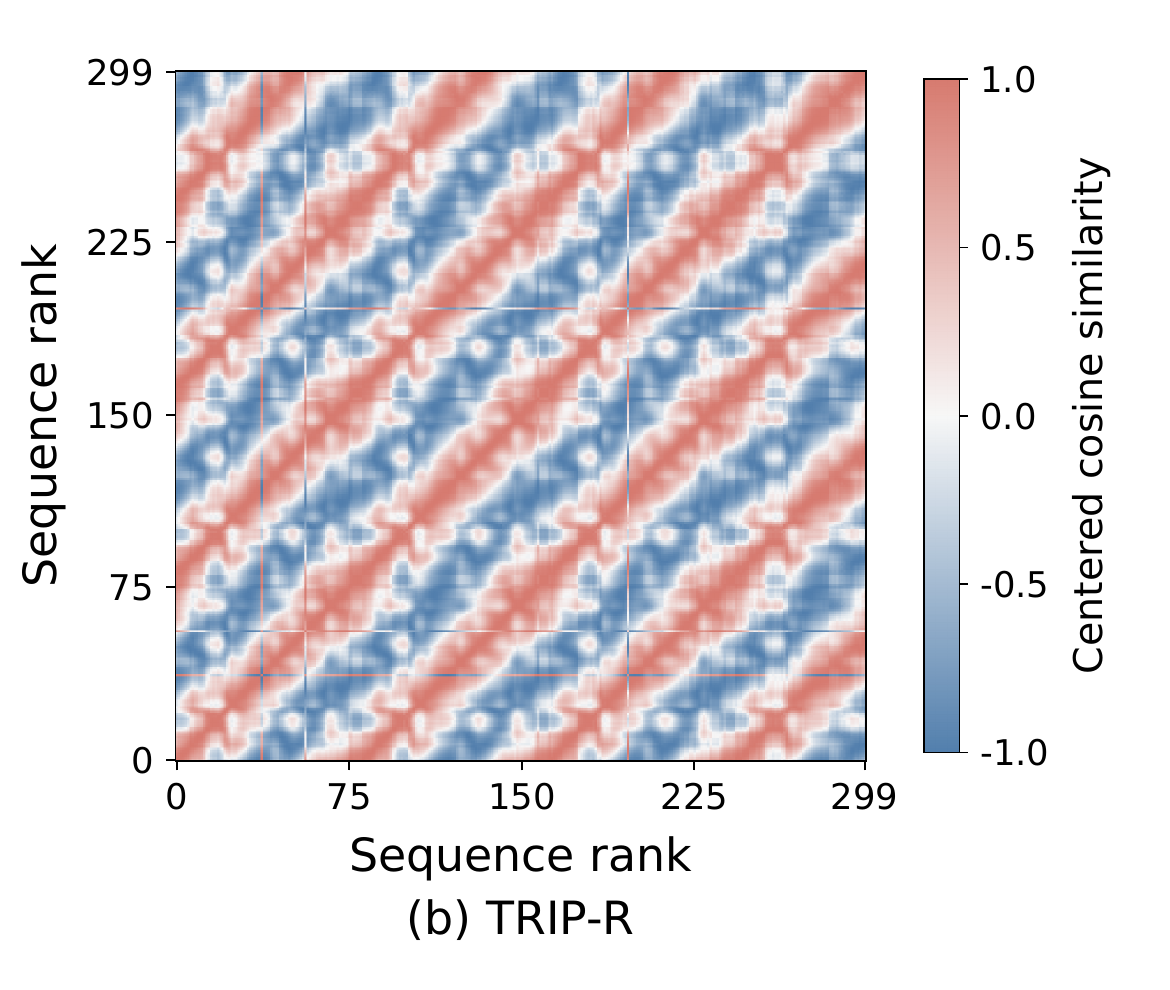}
  \caption{Centered cosine similarity of learned FRE representations on TaobaoAd: (a) TRIP-B and (b) TRIP-R.}
  \Description{Two heatmaps compare learned rank representations over 300 sequence ranks. TRIP-B shows nonuniform blocks of positive and negative similarity. TRIP-R shows more regular repeated diagonal bands. Both color scales range from minus one to one.}
  \label{fig:learned-rank-similarity}
\end{figure}

\begin{figure*}[t]
  \centering
  \includegraphics[width=0.86\textwidth]{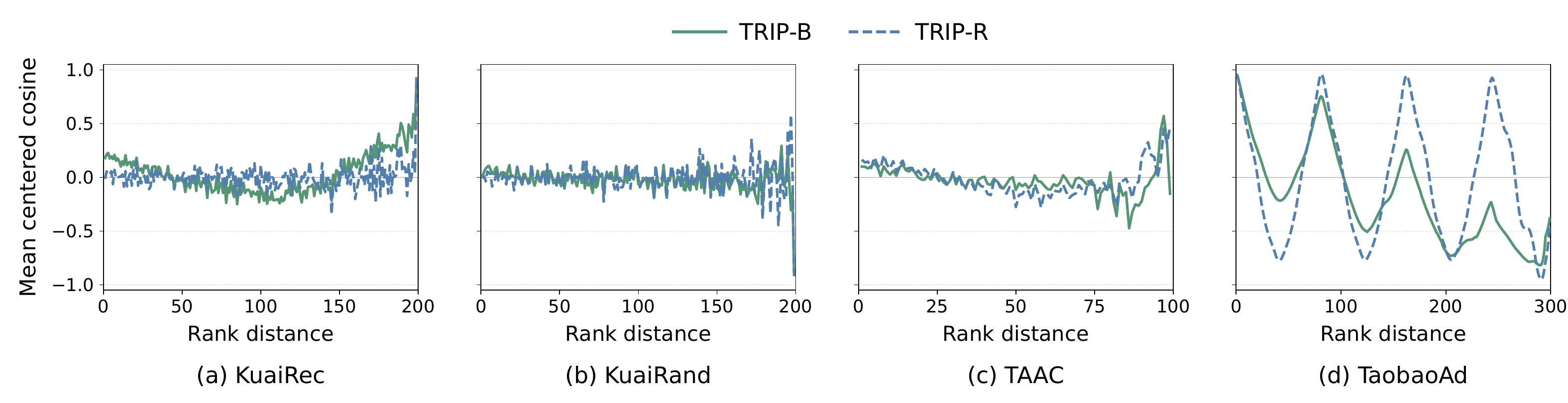}
  \caption{Mean centered FRE cosine similarity versus rank distance (solid: TRIP-B; dashed: TRIP-R).}
  \Description{Four panels compare rank-distance similarity curves. TaobaoAd shows pronounced oscillations, particularly for TRIP-R; the other datasets exhibit weaker oscillations over most distances.}
  \label{fig:rank-lag-similarity}
  \vspace{4pt}
  \includegraphics[width=0.86\textwidth]{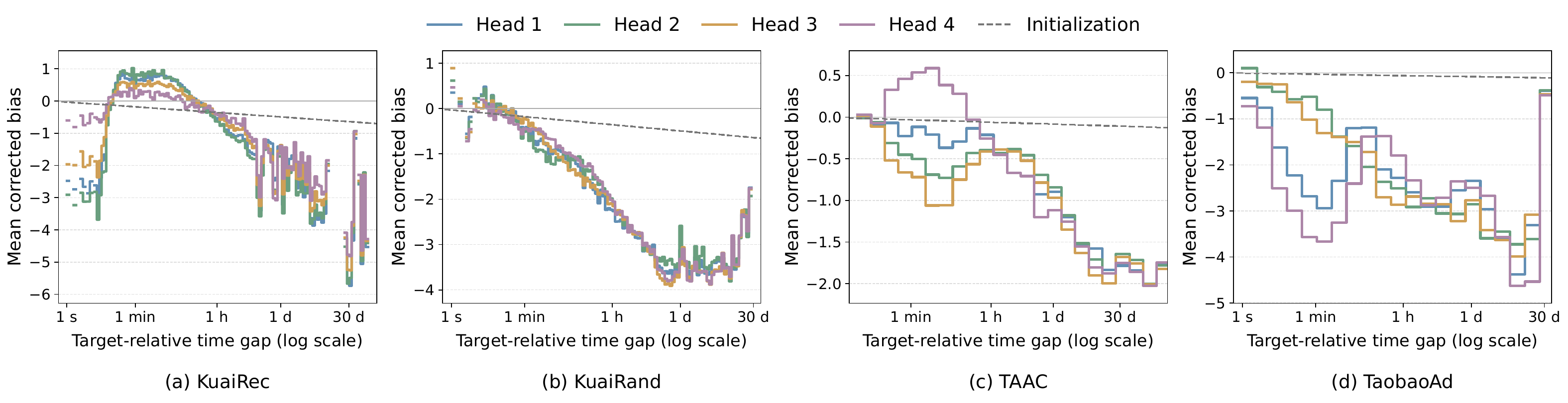}
  \caption{Mean density-corrected temporal biases $\beta_{h,i}$ (dashed: uncorrected initialization).}
  \Description{Four panels show learned temporal potentials after sample-wise interval-occupancy correction, with target-relative elapsed time on logarithmic horizontal axes.}
  \label{fig:density-corrected-temporal-biases}
\end{figure*}

\subsection{Module Implementation}
\label{app:module-details}
\label{app:temporal-details}
\label{app:tgci-details}

\stitle{Encoding Initialization.}
FRE initializes $\boldsymbol\mu$ to zero, $A$ from a normal distribution with standard deviation $1/\sqrt K$, and $B$ from a truncated normal distribution with standard deviation 0.02, without PCA initialization. TRIP-B uses $j_i=\min(J_{\max},1+\lfloor m_{\mathrm B}\log_2(1+\delta_i)\rfloor)$ to index a learnable embedding table $\Theta\in\mathbb R^{(J_{\max}+1)\times H}$, with $\phi_h(j_i;\theta_h)=\Theta_{j_i,h}$. Each head shares its temporal potentials across prediction-side queries. The table is initialized as $\Theta_{j,h}=-0.005j$, and occupancy is counted over valid events before compression. TRIP-R uses fixed RoPE frequencies with base $10^4$.

\stitle{Reconstruction and Summary Details.}
Temporal groups with at least two events retain both masked and visible items. The reconstruction weight increases linearly to $\lambda_{\mathrm{rec}}$ over the first 5\% of training steps. FRE is not reapplied after summary compression. For TRIP-R, each summary's representative timestamp is the rounded mean of its constituent event timestamps, from which target-relative coordinates are recomputed for subsequent blocks.

\section{Supplementary Experiments}
\label{app:supplementary-experiments}

\begin{figure*}[t]
  \centering
  \includegraphics[width=0.95\textwidth]{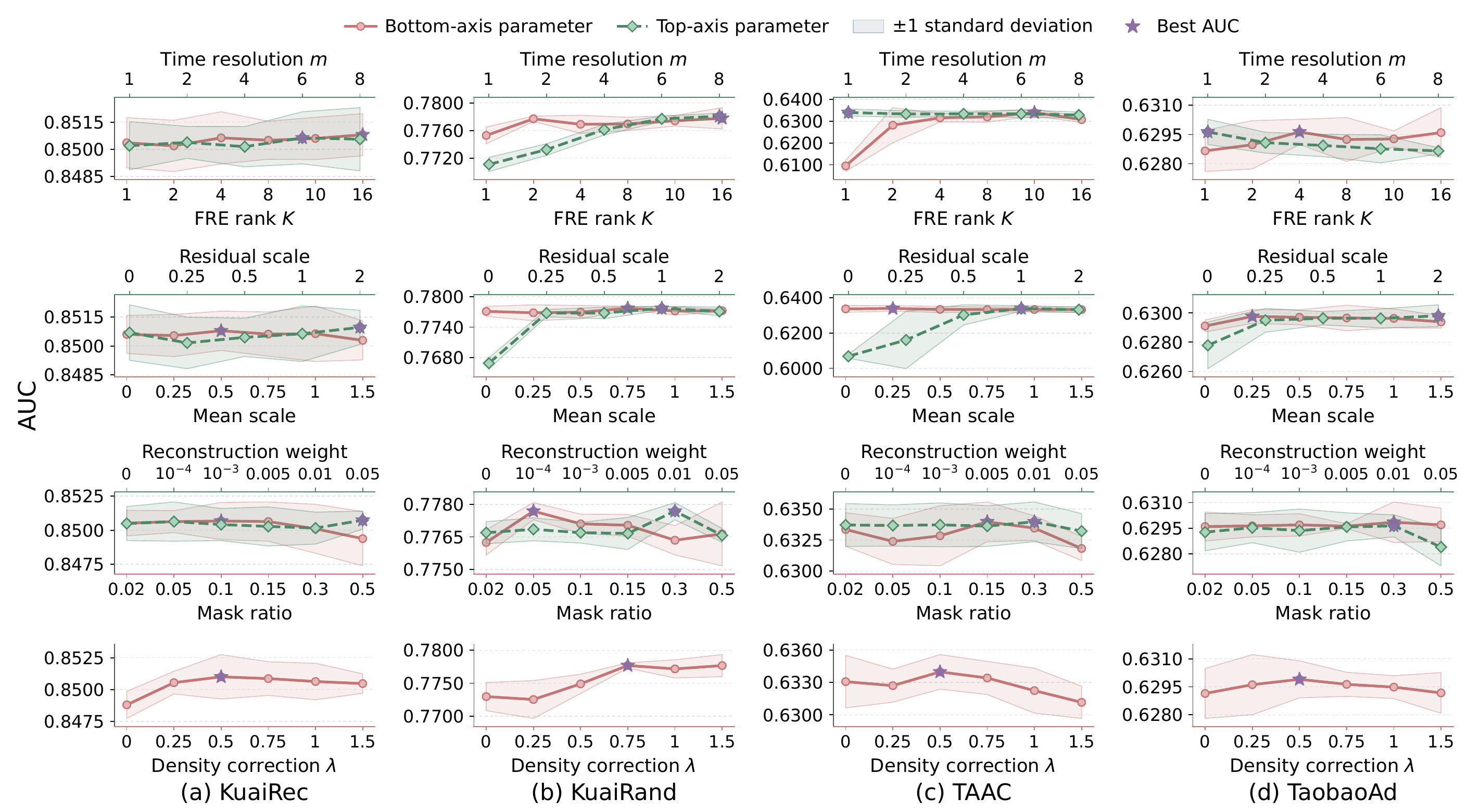}
  \caption{One-factor sensitivity of TRIP-B on four datasets. The first three rows pair independent parameter sweeps on bottom (solid) and top (dashed) axes; the last row varies density-correction strength. Curves show mean test AUC over three seeds.}
  \Description{A four-row, four-column grid covers seven parameters. Rows pair FRE rank K with time resolution m, mean with residual scale, and masking ratio with reconstruction weight; the last row shows density-correction strength. Red solid curves use bottom axes and green dashed curves use top axes. Each curve is a separate one-factor sweep, not a joint variation of both parameters.}
  \label{fig:parameter-sensitivity}
\end{figure*}

\subsection{Additional Ablation Results}
\label{app:complete-main-results}
\label{app:ablation-details}
\label{app:rank-design}
Tables~\ref{tab:main_results} and~\ref{tab:rank-encoding-replacements} report overall performance and rank replacements in the main text. Here we specify the variants for ablation analysis.

\stitle{Variant Definitions.}
TGCA removal restores event-level causal attention while retaining groups, summaries, and reconstruction. Reconstruction removal retains masking; the joint removal disables both. Target-awareness removal changes only the prediction-time reference while keeping all other components unchanged. All ablated variants are compared with the full TRIP-B model.

\stitle{Grouping Controls.}
\label{app:grouping-details}
For the comparison in Figure~\ref{fig:grouping-comparison}, both alternatives begin with the original temporal grouping and retain each group label's event count. Random permutes group labels among valid events using a deterministic fixed-seed hash, so an individual sample's assignment remains unchanged across training steps. Rank-contiguous reorders the original group sizes using a fixed permutation of group labels, then partitions consecutive valid ranks according to those sizes. Its segments are neither equal-sized nor fixed-width; all strategies retain the same number of group summaries.

\subsection{Learned Positional Representations}
\label{app:representation-analysis}
\label{sec:learned-rank-representations}
\label{app:temporal-analysis}
\label{sec:learned-temporal-potentials}
\stitle{Learned Rank Geometry.}
Figures~\ref{fig:learned-rank-similarity} and~\ref{fig:rank-lag-similarity} show that FRE does not impose a uniform distance-decay geometry. On TaobaoAd, TRIP-B learns nonuniform rank relationships, whereas TRIP-R exhibits more regular diagonal bands and pronounced rank-distance oscillations. The other datasets show weaker oscillatory structure. These differences illustrate how low-rank sharing accommodates dataset-specific positional relationships and rank representations adapted to the temporal formulation.

\stitle{Temporal Preference and Density Correction.}
Figure~\ref{fig:density-corrected-temporal-biases} reports biases averaged over events within each interval using 8,192 test samples per dataset. Compared with the raw potentials in Figure~\ref{fig:learned-temporal-potentials}, occupancy correction changes the relative prominence of temporal windows, showing that learned preference alone does not determine the applied bias. Since the same interval penalty is subtracted across heads, correction preserves their preference differences while reducing event-multiplicity effects. Thus, TRTE separates head-specific temporal preference from sample-specific history density.

\subsection{Efficiency Analysis}
\label{app:efficiency-details}
On TaobaoAd, TRIP-B takes 72.63 ms/step, compared with 73.78 for OneTrans, 147.23 for HyFormer, 124.20 for MixFormer, and 123.88 for TokenFormer. This measurement uses an optimized temporal-bias implementation; Table~\ref{tab:summary-efficiency} compares runs with and without compression under the same standard implementation, so its relative savings are unaffected. TRIP retains OneTrans's dense-attention formulation: its full-history first block adds positional-encoding and reconstruction overhead, offset by shorter group-summary sequences in later blocks. For $N_{\mathrm{blk}}$ blocks, the dense attention cost is $O((L+N_{\mathrm{ns}})^2D)+(N_{\mathrm{blk}}-1)O((G+N_{\mathrm{ns}})^2D)$. For candidates sharing a request timestamp and candidate-independent historical inputs, history-side states and key/value projections can in principle be reused, since historical tokens do not attend to NS tokens.

\subsection{Hyperparameter Analysis}
\label{app:parameter-sensitivity}
\label{app:additional-parameter-sensitivity}
\label{sec:parameter-sensitivity}
Figure~\ref{fig:parameter-sensitivity} examines seven TRIP-B hyperparameters across four datasets, varying one at a time with the others fixed. Each sweep, including its reference point, is independently rerun with three seeds; reference means may therefore differ across sweeps and from Table~\ref{tab:main_results}. Validation-selected settings in Table~\ref{tab:trip-hyperparameters} need not coincide with these test-set peaks. Branch-scale axes retain the original configuration values.

\stitle{Factorization Width.}
FRE benefits from sufficient rank-specific capacity, but larger $K$ is not uniformly better. On TAAC, mean AUC rises from 0.6095 at $K=1$ to 0.6340 at $K=10$, then declines at $K=16$. KuaiRec is much less sensitive, with AUC ranging from 0.8502 to 0.8508 across the tested widths; on KuaiRand, $K=2$ already approaches the best result. These differences favor a compact, dataset-adaptive factorization: shared basis vectors capture useful ordinal structure, while the required capacity for distinguishing ranks varies across histories.

\stitle{Temporal Resolution.}
Finer target-relative coordinates improve temporal discrimination on KuaiRand: increasing $m$ from 1 to 8 raises mean AUC from 0.7711 to 0.7781. In contrast, TAAC and TaobaoAd attain their best results at $m=1$, and KuaiRec shows only small fluctuations. Thus, increasing resolution is useful when finer distinctions help history reading, but does not yield universal gains. Here, $m$ subdivides TRTE's logarithmic time scale without changing TGCI's temporal grouping, separating the granularity of temporal reading from that of group-level interaction.

\stitle{Density-Correction Strength.}
All four datasets achieve their highest observed AUC with nonzero $\lambda_{\mathrm{cnt}}$. The peaks occur at 0.5 on KuaiRec, TAAC, and TaobaoAd, and at 0.75 on KuaiRand. For example, KuaiRec improves from 0.8488 without correction to 0.8510 at 0.5. Moderate correction counteracts the attention-mass advantage of densely populated intervals, allowing temporal preference and content relevance to guide reading. Further increasing the penalty brings no consistent benefit, indicating that occupancy should be corrected without excessively suppressing dense but potentially informative history.

\stitle{FRE Branch Scales.}
The shared-offset scale has a comparatively small effect: the AUC range across its tested values is below 0.001 on every dataset. Rank-specific variation is more consequential, particularly on TAAC and KuaiRand. Setting the configured rank-branch scale to zero yields AUCs of 0.6068 and 0.7669, compared with 0.6340 and 0.7777 at configuration scale one. Unlike a shared offset, the rank-specific branch distinguishes ordinal contexts. These results support FRE's separation of shared structure and rank-dependent variation, with the latter providing the essential positional distinctions on these datasets.

\stitle{Masking and Reconstruction.}
The masking-ratio sweeps peak at $\rho=0.1$, 0.05, 0.15, and 0.3 on KuaiRec, KuaiRand, TAAC, and TaobaoAd, respectively. Raising $\rho$ to 0.5 reduces AUC relative to these peaks, particularly on KuaiRec and TAAC. Masking encourages contextual recovery, but excessive removal leaves less item content for both reconstruction and CTR prediction. The reconstruction-weight sweeps similarly favor balanced supervision: $\lambda_{\mathrm{rec}}=0.01$ gives the highest AUC on KuaiRand, TAAC, and TaobaoAd, whereas KuaiRec peaks at 0.05. Increasing the weight to 0.05 reduces AUC on the former three datasets. Together, these patterns favor auxiliary supervision that strengthens contextual learning without overwhelming the prediction objective.

\subsection{KuaiRand Probability Calibration}
\label{app:kuairand-calibration}
Probability diagnostics on KuaiRand (seed 42) show that TRIP-B predicts a mean positive-feedback probability of 46.74\%, above the observed rate of 37.67\%. This bias is particularly pronounced in \texttt{tab=0}, which contains 19.03\% of test samples: its observed positive rate is 11.64\%, compared with a mean prediction of 36.05\%. This scenario contributes 69.93\% of the sample-weighted net LogLoss gap relative to MixFormer. Such probability overestimation increases the loss on negative examples, despite improved ranking in the dominant scenario. Positive-slope affine logit calibration fitted only on validation data reduces TRIP-B's LogLoss from 0.5778 to 0.5471 without changing AUC. Applying validation-fitted calibration to both models reduces TRIP-B's LogLoss gap to MixFormer from 0.0300 to 0.0011 while preserving its AUC advantage. This suggests that the higher uncalibrated LogLoss largely reflects correctable probability miscalibration rather than weaker ranking discrimination.

\end{document}